\documentclass[aps,prx,twocolumn,amsmath,amssymb,showpacs,superscriptaddress,longbibliography]{revtex4-1}

\usepackage[utf8]{inputenc}
\usepackage{graphicx}
\usepackage{amsmath,amsthm,amssymb}
\usepackage[normalem]{ulem}
\usepackage[table]{xcolor}
\usepackage{xr}
\usepackage{stmaryrd}
\usepackage[english]{babel}
\usepackage{hyperref}
\graphicspath{{figures}}
\newcommand*\diff{\mathop{}\!\mathrm{d}}

\usepackage{booktabs}
\usepackage{array}
\usepackage{xcolor,colortbl}
\definecolor{greyhead}{gray}{0.83}
\definecolor{greyrow}{gray}{0.96}

\begin{document}

\title{Inferring Spatial Source of Disease Outbreaks using Maximum Entropy}
\author{Mehrad Ansari}
\affiliation{Department of Chemical Engineering, University of Rochester, Rochester, NY, 14627, USA.}
\author{David Soriano-Pa\~nos}
\affiliation{Department of Condensed Matter Physics and Institute for Biocomputation and Physics of Complex Systems, University of Zaragoza, E-50009 Zaragoza, Spain.}
\author{Gourab Ghoshal}
\affiliation{Department of Physics \& Astronomy and Computer Science, University of Rochester, Rochester, NY, 14627, USA.}
\author{Andrew D. White}
\thanks{andrew.white@rochester.edu}
\affiliation{Department of Chemical Engineering, University of Rochester, Rochester, NY, 14627, USA.}

\begin{abstract}
     Mathematical modeling of disease outbreaks can infer the future trajectory of an epidemic, which can inform policy decisions.
     Another task is inferring the origin of a disease, which is relatively difficult with current mathematical models.
     Such frameworks---across varying levels of complexity---are typically sensitive to input data on epidemic parameters, case-counts and mortality rates, which are generally noisy and incomplete.
     To alleviate these limitations, we propose a maximum entropy framework that fits epidemiological models, provides a calibrated infection origin probabilities, and is robust to noise due to a prior belief model.
     Maximum entropy is agnostic to the parameters or model structure used and allows for flexible use when faced with sparse data conditions and incomplete knowledge in the dynamical phase of disease-spread, providing for more reliable modeling at early stages of outbreaks. 
     We evaluate the performance of our model by predicting future disease trajectories in synthetic graph networks and the real mobility network of New York state.
     In addition, unlike existing approaches, we demonstrate that the method can be used to infer the origin of the outbreak with accurate confidence.
     Indeed, despite the prevalent belief on the feasibility of contact-tracing being limited to the initial stages of an outbreak, we report the possibility of reconstructing early disease dynamics, including the epidemic seed, at advanced stages.
\end{abstract}

\maketitle

\section{Introduction}
The spread of SARS-CoV-2 virus constitutes the most recent example of the vulnerability of modern society  to the spread of communicable diseases~\cite{Estrada2020,world2020coronavirus, Hazarie_2021}.
In particular, the combination of features such as extensive trans- and intra-national transportation networks, shortening travel-time between faraway regions \cite{candido2020routes,chinazzi2020effect,gilbert2020preparedness}, the existence of important socioeconomic inequities~\cite{abedi2021racial,ahmed2020inequality, Barbosa_2021} and the phenomenon of rapid urbanization~\cite{berry2008urbanization,bertinelli2004urbanization} have conspired to give rise to the unprecedented speed at which SARS-CoV-2 has advanced, becoming a global threat within a few months of the (reported) initial outbreak.  

The risk of significant harm to society from an epidemic is increased when there is an initial lack of knowledge about the epidemiological features of a novel pathogen, limiting the use effective of medical treatments or vaccines to slow down progression at the early stages of the outbreak. 
Indeed, early attempts at mitigation resorted to non-pharmaceutical interventions such as recommending hand-washing, hygienic measures, social distancing, travel restrictions, and population confinement via stay-at-home orders~\cite{kraemer2020, maier2020,gatto2020}.
A key tool for devising and assessing the effectiveness of such measures is mathematical modeling of the epidemic trajectories under various scenarios. The advantage of such models are two-fold: on the one hand, epidemic models provide short-term forecasts on the evolution of an outbreak, providing useful information to assess the potential harmfulness of the pathogen and act accordingly to reduce their impact.
On the other hand, the different layers of complexity introduced in the epidemic models has boosted their use as benchmarks to devise cost-effective non-pharmaceutical interventions aimed at hindering the spread of the disease~\cite{aleta2020, dehning2020}. 

Regardless of their stochastic or deterministic nature~\cite{anderson1992infectious, keeling2011modeling, vynnycky2010introduction}, the successful application of epidemic models to provide reliable forecasts is tightly linked with the correct estimation of their relevant parameters.
Early on in an epidemic, the key parameters describing the spread of the infection are highly uncertain and this uncertainty can severely impact the predicted outcomes~\cite{meehan2020modelling}.
This becomes particularly relevant in the context of highly complex compartmental models that produce wildly-varying degenerate trajectories in the short-term dynamics, even for small changes in the parameter-estimates~\cite{roosa2019assessing,castro2020turning}.
While, this degeneracy dissipates in the long-term dynamics due to exponential growth encoded in the equations, even minor inaccuracies in the epidemic parameters limits reliable predictions to at most a few weeks in the future~\cite{wilke2020predicting}.
Given this, the practical efficacy of epidemiological models is in providing a range of possible outcomes, rather than producing precise quantitative predictions \cite{jewell2020predictive}.

\begin{figure*}[tb!]
    \centering
    \includegraphics[width=\textwidth]{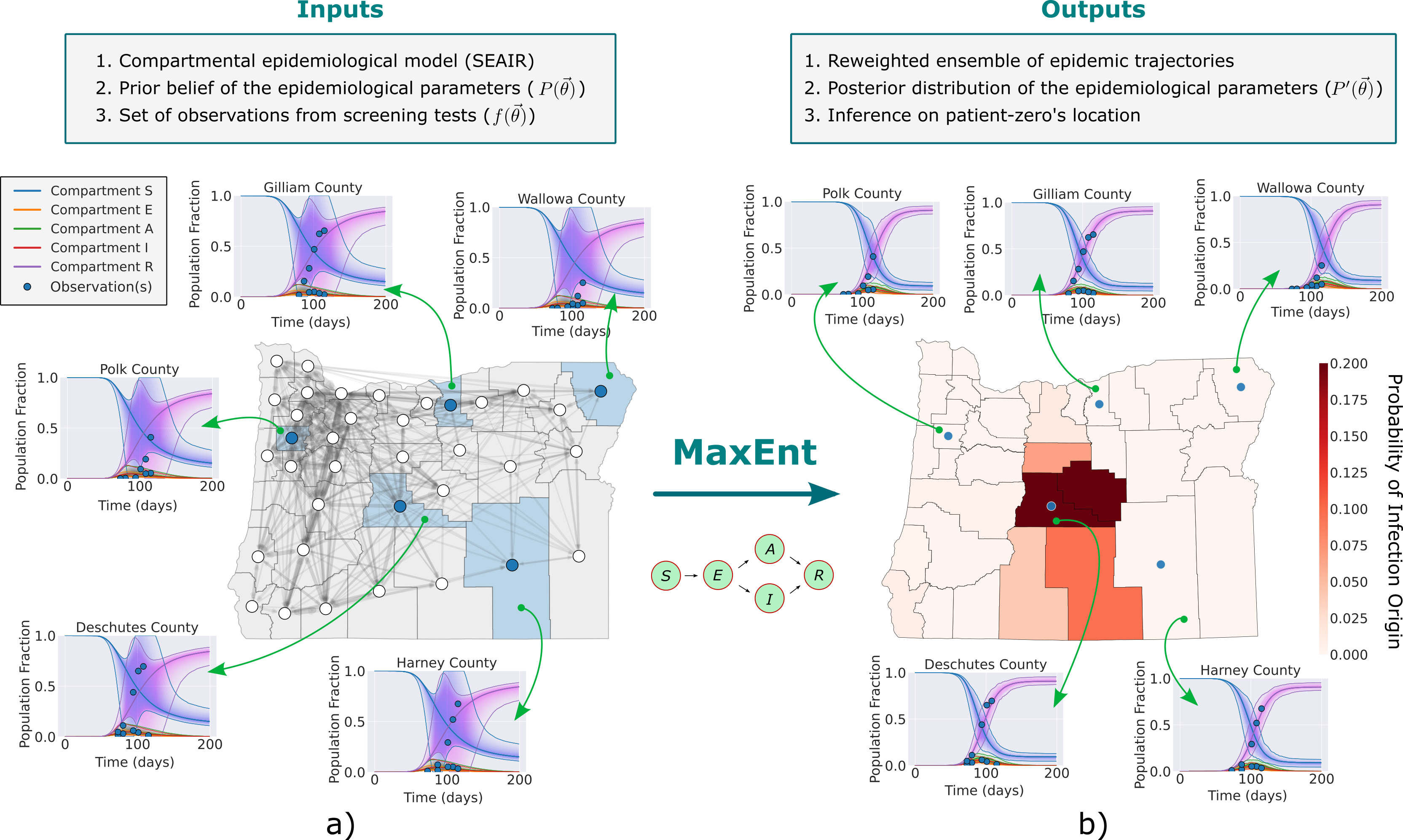}
    \caption{{\bf High-level model overview}. a) Model inputs: an {\bf SEAIR} compartmental epidemiological model, prior belief of the epidemiological parameters, and a set of sparse observations that come from disease screening tests. The contact network in a metapopulation can be represented as a network graph. The infection starts at an unknown origin and spreads through the network. We generate a large set of trajectories and explore the epidemic trajectory space over a high variance prior belief for the epidemiology parameters. The large variance is represented as the shaded areas with 80\% confidence intervals. The infections starts in a single node in each trajectory series but that node varies over the next trajectories. b) Model outputs: MaxEnt re-weighted ensemble of trajectories given the observations, posterior distributions of the parameters and predicted infection origin. The re-weighted trajectories allow us to predict how the disease spreads through the network and infer the location for the source of infection.}  \label{fig:concept}
\end{figure*}
Multiple ways to infer epidemiological parameters have been proposed in the literature.
One typical method is to use maximum likelihood approaches, where parameter values are chosen to maximize the likelihood of observing the experimentally-measured data (observations), given some prior distribution on the parameters~\cite{Pandey2013DengueSIR, Lavielle2011HIVSIR}.
A disadvantage of this method is that the functional form of the likelihood function must be known or approximated to perform maximization. Another approach is least-squares fitting, which employs various optimization methods, including but not limited to: Markov chain Monte Carlo~\cite{New2009SIRMCMC,Morton2005SIRMCMC,Cauchemez2008SIRMCMC,Talawar2016SIRMCMC}, sequential Monte Carlo~\cite{He2010SIRSMC,King2008SIRSMC,Ionides2006SIRSMC}, trajectory matching~\cite{bock1983TrajectoryMatching,Arora2004TrajectoryMatching,Banks1981TrajectoryMatching,Ciupe2006TrajectoryMatching,Biegler1986TrajectoryMatching} and machine learning methods like support vector machines~\cite{Yang2013LeastSquaresSVM}.
Other approaches include generalized profiling~\cite{Hooker2011SIRGeneralizedProfiling}, approximate Bayesian computation~\cite{Blum2010SIRABC,Toni2009SIRABC,Kypraios2017SIRABC}, derivative-free optimization~\cite{Rogalsky2012SIRGradientFree,Iacoviello2008SIRGradientFree} and Bayesian inference~\cite{Streftaris2004SIRBayesianInference,Clancy2008SIRBayesianInference,DEMIRIS2005SIRBayesianInference,Altarelli2014SIRBayesianInference,ElMaroufy2016SIRBayesianInference}.
Furthermore, most of the epidemiological models in the literature focus on forward dynamics of the diffusion of the pathogen through the network, while the backward-dynamics problem of identifying the diffusion source has been comparatively less studied~\cite{shah2020finding, shen2016locating, pinto2012locating}.
Such an analysis bears significant importance in guiding systematic contact-tracing and increasing the chance of early containment of an outbreak.

An approach that circumvents these difficulties is a well-known method from statistical mechanics, maximum entropy (MaxEnt) biasing.
MaxEnt has been proven to be successful in various settings such as molecular dynamics simulations~\cite{marinelli2015ensemble, cesari2018using, Amirkulova2019}, ecology~\cite{shipley2006plant,harte2008maximum,dewar2008statistical,favretti2018remarks}, nuclear magnetic resonance spectroscopy~\cite{sibisi1984maximum, hoch2014nonuniform}, x-ray diffraction~\cite{kitaura2002formation, andersen2014location}, electron microscopy~\cite{ferrige1992application, kimoto2010local}, economics~\cite{scharfenaker2020maximum} and neuroscience~\cite{schneidman2006weak, tang2008maximum, granot2013stimulus, watanabe2013pairwise}.
This method uses the principle of entropy to measure the difference between two distributions or trajectories and applies a change using Lagrange multipliers to alter a given distribution to match a target one, while maximizing the entropy (and thus, effecting minimal change)~\cite{barrett2021simulation}.
This approach is highly promising in the context of epidemic modeling, as it mitigates the need for designing complex compartmental models and having to make a lot of simplifying assumptions.
As remarked in \cite{vespignani2020modelling}: \textit{``What has been produced the day before often must be completely revised the day after because a new piece of information has arrived''}.
This approach relies more on daily (weekly) evidence, rather than relying on uncertain early estimates of disease parameters, especially at the early stages of an epidemic outbreak.
A few instances of applying MaxEnt to characterize epidemic spreading exist in the literature.
In~\cite{artalejo2011sis} MaxEnt is used to bias the epidemic curves generated by mean-field {\bf SIS} and {\bf SIR} compartmental models to reproduce a set of empirical observations and uncover probability distributions used for contagion and recovery events.
~\citet{harding2020population} propose a MaxEnt approach to modify a {\bf SIS} framework running on a contact network to model the time-varying nature of human mobility in response to the diffusion of an epidemic outbreak.

Here, we explore the use of MaxEnt biasing when more layers of complexity are added to the dynamic equations governing the advance of an epidemic.
To do so, we consider a more elaborated compartmental scheme, the {\bf SEAIR} model, running on metapopulations~\cite{hanski1998metapopulation,ball2015seven} to accommodate different realistic features such as human mobility, the relevance of the incubation period of one pathogen or the existence of asymptomatic infectious individuals~\cite{Estrada2020}.
We show that MaxEnt biasing allows for both predicting future trajectories as well as inferring the source of infection.
In Fig.~\ref{fig:concept} we represent a high-level overview of the framework.
Graphs in this work were generated using NetworkX~\cite{hagberg2008exploring}.
Model inputs include a compartmental epidemiology model, prior belief for its parameters and a set of sparse observations.
The prior belief on the model parameters can include a relatively large variance, making our approach highly applicable to risk assessment analysis at the early stages of the outbreak, where the true parameters are unknown.
The observations are weekly average data obtained by disease test screenings that contain random noise.
This noise accounts for the uncertainty associated with the number of infected individuals due to the variance of testing policies across a metapopulation. The output is the MaxEnt re-weighted trajectories that are used for inference on the epidemic spread and the source of infection.
Using this method applies minimal change to the model's original output, without altering the parameters directly. The premise of this change is that the original model is treated as well-trusted but only slightly incorrect, with the intent of improving predictive accuracy for future events by matching the model's output to experimental data (observations).
However, experimental data is known to contain systematic error, so we include a formulation of MaxEnt that accounts for some bias.
This method is agnostic to the functional form of the original model; given that it re-weights paths produced by sampling model parameters, which can be done a priori, it can be treated as a black box.
This also has the advantage that the method's computational complexity scales with only the number of paths sampled and number of target functions, rather than the number of model parameters~\cite{barrett2021simulation}.

The manuscript is organized as follows. In Sec.~\ref{sec:maxent}, we describe the theory of MaxEnt applied to a general model function, $P(\vec{\theta})$ with parameters $\vec{\theta}$, and describe the procedure for MaxEnt path biasing.
In Sec.~\ref{sec:compartment} we describe the underlying equations of the {\bf SEAIR} model occurring on a metapopulation framework.
In Sec.~\ref{sec:results} we present results on both synthetic and real-world metapopulation mobility networks and demonstrate how the method can predict infection spread, make a high certainty inference on the source of an epidemic using the posterior re-weighted trajectory from the MaxEnt approach.
In particular, we demonstrate that this inference can be done even in late stages of the disease dynamics.
In Sec.~\ref{sec:discuss} we end with a discussion of the implications of our findings. 

\section{Theory}
\subsection{Maximum Entropy with Uncertainty}
\label{sec:maxent}
Consider for a given simulator $f(\vec{\theta})$ with a set of parameters $\vec{\theta}$, we have a prior distribution of parameters $P(\vec{\theta})$.
For example, the function $f(\vec{\theta})$ can be a system of ODEs in a compartmental epidemiology model.
Given a set of $N$ observations with uncertainty $\epsilon_k$, where $\{\bar{g}\}_k$, $k\in[1,\ldots,N]$, we constrain our prior model $P(\vec{\theta})$ such that:
\begin{equation}
    \int\diff\vec{\theta}\diff\vec{\epsilon} P'(\vec{
    \theta}) (g_k[f(\vec{\theta})] + \epsilon_k) = E[g_k + \epsilon_k] = \bar{g}_k \forall k
\end{equation}
This means that we want the average over the posterior distribution $P'(\vec{\theta})$ to match the observation data with some allowable disagreement based on $\{\epsilon_k\}$.
Note that unlike in Bayesian frameworks, the mentioned average disagreement with the data is optional (i.e $P_0(\epsilon_k) = \delta(\epsilon_k = 0)$).
However, in our settings, the Laplace distribution prior $P_0(\epsilon)$ is used to account for this error with a given standard deviation $\sigma_0$, thus:
\begin{equation}
    \label{eq:Laplace_prior}
    P_0(\epsilon) = \frac{e^{\frac{-\epsilon^2}{2\sigma_0^2}}}{\sqrt{2\pi}\sigma_0}
\end{equation}
The posterior distribution $P'(\theta)$ that satisfies $N$ constraints is given by~\cite{Roux2013statistical,Pitera2012,Amirkulova2019,cesari2016MaxEntUncertainty}:
\begin{eqnarray}
  P'(\vec{\theta}, \vec{\epsilon}) &= \frac{1}{Z'}P(\vec{\theta})\prod_k^Ne^{-\lambda_kg_k[f(\vec{\theta})]}e^{-\lambda_k\epsilon_k}P_0(\epsilon_k), \\
  Z' &= \int\diff \vec{\theta} \diff \vec{\epsilon} P(\vec{\theta}) P_0(\epsilon) e^{-\sum_k\lambda_k (g[f(\vec{\theta})] + \epsilon_k)},
\end{eqnarray}
where $Z'$ is a normalization constant and $\lambda_k$ values are iteratively updated using gradient descent to satisfy the constraint $E[g_k + \epsilon_k] = \bar{g}_k$.
The MaxEnt framework suggests a strong belief in our prior distribution of parameters in this setting, which reflects the use of approximately correct parameters.
Consider  health emergencies like COVID-19 global pandemic.
At the initial phase of the outbreak, little to no information is available on the pathogen, its transmissibility and the general parameters that describe how the infection spreads.
However, one can make an educated guess for the average values of these parameters and make reliable predictions by taking advantage of the ensemble of outcomes from MaxEnt, whose means agree with observed data.
In this setting, the observations can be the number of confirmed disease, given some random noise to account for uncertainty.
More information on the MaxEnt model implemented in this study can be found in the work of \citet{barrett2021simulation}.

\subsection{Epidemic Model}
\label{sec:compartment}
Epidemic spreading can be represented as a reaction-diffusion process where the reaction term refers to the contagion events triggered by the interaction between infected and susceptible hosts whereas the diffusion phase corresponds to the spatial dissemination of the population across the system under study.
In this sense, metapopulations, originally introduced in the field of ecology, represent a convenient framework, balancing complexity with analytical tractability, to account for the impact of mobility on epidemic spreading~\cite{grenfell1997meta,watts2005multiscale,colizza2007reaction}.
Metapopulations are comprised of spatial patches (nodes) where local populations interact in a mean-field manner,  connected via flows (edges) corresponding to movement of individuals between patches.
The spatial resolution of the spatial patch may vary (neighborhoods, zip-codes, districts, cities etc.) depending upon the granularity of the input data, or the scale at which the dynamics are being modeled. 
In what is to follow, we assume that our metapopulation is composed of $N_P$ patches and that each patch $i$ is populated by $n_i$ residents.

\begin{figure}[b!]
    \centering
    \includegraphics[width=0.4\textwidth]{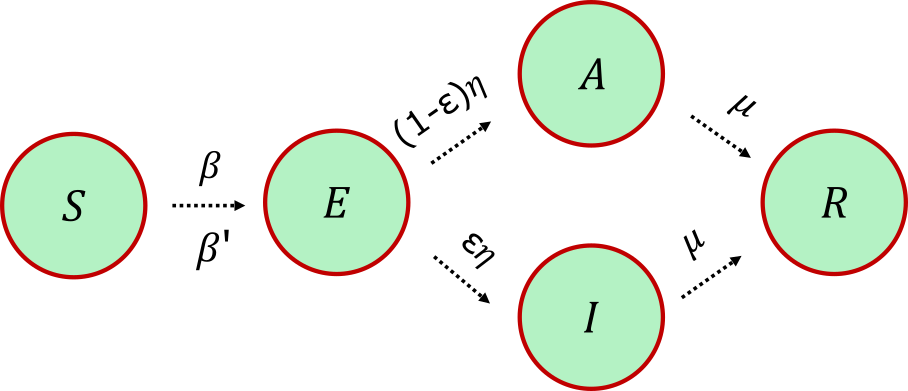}
    \caption{{\bf SEAIR compartmental scheme.} Populations in each patch can be any of Susceptible , Exposed , Asymptomatic, Infected  and Resolved. Susceptible ({\bf S}) individuals can get exposed ({\bf E}) to the disease through I-S and A-S contacts with infectivity rates $\beta$ and $\beta'$. Once exposed, they become asymptomatic ({\bf A}) or infected ({\bf I}) at rate $\eta$. They finally recover or die at rate $\mu$ and become resolved ({\bf R}). $\epsilon$ accounts for the fraction of infected (symptomatic) individuals. \label{fig:SEAIR}}
\end{figure}

To model the disease spread, we consider a variant of the {\bf S}usceptible-{\bf E}xposed-{\bf I}nfected-{\bf R}emoved ({\bf SEIR}) model to account for the existence of (A)symptomatic individuals.
With the addition of compartment {\bf A}, our model is denoted as the {\bf SEAIR} model. 
The choice for this particular flavor of compartments was inspired by its relevance in modeling the evolution of the current COVID-19 pandemic \cite{zhou2020pneumonia, wu2020new}. The schematic of the model is detailed in Fig.~\ref{fig:SEAIR}.
Susceptible individuals become exposed by having contacts with asymptomatic and infectious agents with probability of $\Pi$.
Let $\beta$ and $\beta'$ be infectivity rates for \textbf{I-S} and \textbf{A-S} contacts, respectively.
Once exposed, susceptible agents turn into asymptomatic or infected at rate $\eta$.
The fraction of infected (symptomatic) individuals is denoted with $\epsilon$.
Finally, they recover or die at escape rate $\mu$ and become resolved.
Note that once resolved, the individuals have lifelong immunity and can no longer be infected.

Considering mobility, we follow the movement-interaction-return scheme introduced in~\cite{gomez2018critical} to reflect the impact of commuting mobility on epidemic spreading.
At the movement stage, the individuals decide whether to move or not with a probability $p$, which is identified as the degree of mobility of the population.
If they move, they choose their destination according to the flows encoded in the links of the metapopulation. Following the redistribution of the population, contagion and recovery processes take place at the interaction stage, modifying the epidemic state of the population accordingly.
Finally, to reflect the recurrent nature of daily human movements, all the agents come back to their associated residential areas.

 The spreading process is represented through a temporally discretized ODE that includes the spatial distribution of the population as well as their mobility patterns~\cite{arenas2020mathematical}.
Here we aim at characterizing the evolution of the fraction of agents in state $m$ (where $m\in\{S,E,A,I,R\})$ associated with each node $i$, denoted in the following by $\rho^m_i(t)$. The temporal evolution of these quantities are given by:
\begin{eqnarray}
\rho^S_i(t+1) &=& (1-\Pi_i(t))\rho^S_i(t) \label{eq:modelS}\\
\rho^E_i(t+1) &=& (1-\eta)\rho^E_i(t) + \rho^S_i(t)\Pi_i(t)\label{eq:modelE}\\
\rho^A_i(t+1) &=& (1-\epsilon)\eta\rho^E_i(t)+(1-\mu)\rho^A_i(t)\ ,\\
\rho^I_i(t+1) &=& \epsilon\eta\rho^E_i(t) + (1-\mu)\rho^I_i(t)\ ,\\
\rho^R_i(t+1) &=& \rho^R_i(t) +\mu \left[\rho^I_i(t)+\rho^A_i (t)\right]\label{eq:modelD}\ .
\end{eqnarray}
$\Pi_i(t)$ denotes the probability that a susceptible agent associated with node $i$ contracts the disease by making contacts with  an asymptomatic or infected individual.
Under our assumptions regarding human mobility, it can be expressed as:
\begin{equation}
\Pi_i(t) = (1-p)P_i(t) + p\sum\limits_{j=1}^{N_P} R_{ij}P_j(t)\ 
\label{EqPi_i}
\end{equation}
The first term in Eq. \ref{EqPi_i} accounts for the probability of contracting the disease within the residential node, while the second term contains the contractions from neighboring nodes. Therefore, note that  $p=1$ corresponds to a scenario where all the agents follow their usual commuting patterns whereas $p=0$ represents a controlled scenario where mobility is fully suppressed and every agent remains in its associated node.
In this work, we work in the uncontrolled scenario and fix $p=1$ throughout the entire manuscript. In this case, the movements are dictated by the entries of the origin-destination (OD) matrix ${\bf R}$, whose elements $R_{ij}$ denote the probability for one individual residing in patch $i$ moving to $j$.
Assuming that the number of trips recorded between both locations in a real dataset is given by $T_{ij}$, these probabilities are easily computed as $R_{ij} = T_{ij}/\sum\limits_j T_{ij}$\ . 
Likewise, $P_i(t)$ is the probability of getting the disease in node $i$ at time $t$ and $p$ accounts for the degree of mobility of individuals. Under the well-mixed assumption, $P_i(t)$ is written as:
\begin{equation}
P_i (t) = 1-\prod\limits_{j=1}^{N_P} (1-\beta)^{z f(\frac{n_i^{eff}}{a_i})\frac{n^I_{j\rightarrow i}(t)}{n_i^{eff}}}(1-\beta ')^{z f(\frac{n_i^{eff}}{a_i})\frac{n^A_{j\rightarrow i}(t)}{n_i^{eff}}}\ ,
\label{eq:P_i}
\end{equation}


where $n^m_{j\rightarrow i}$ is the number of infectious agents going from $j$ to $i$ belonging to the compartment $m$ and $a_i$ denotes the area of node $i$. In turn, $n_i^{eff}$ encodes the effective population of patch $i$ after population movements.  In particular:
\begin{eqnarray}
n^A_{j\rightarrow i}(t) &=& n_j\rho^A_j(t)\left[(1-p)\delta_{ij} + pR_{ji}\right]\  \\
n^I_{j\rightarrow i}(t) &=&  n_j\rho^I_j(t)\left[(1-p)\delta_{ij} + pR_{ji}\right] \ \\
n_i^{eff} &=& \sum\limits_{j=1}^{N_P} n_j\left[(1-p)\delta_{ij} + pR_{ji}\right]\,\label{eq:Nieff} 
\end{eqnarray}
Note that the product in Eq. \ref{eq:P_i} accounts for the probability for an individual not getting infected while staying in node $i$ and the exponent represents the number of contacts made with the infectious individuals from compartments {\bf A} and {\bf I}. Function $f$ accounts for the dependence of the number of contacts on the population density ($x$) of each node. Our choice for this function is:
\begin{equation}
f(x) = 2- e^{-\xi x}\, .
\end{equation}
where $\xi$ is a constant, accounting for how the number of contacts depend on the population density of one area.
Throughout the manuscript, we fix $\xi = 5\cdot 10^{-3}$ square miles.  Finally, $z$ is a normalization function to ensure that the average number of contacts across the whole population is $\langle k\rangle$. Therefore:
\begin{equation}
z=\frac{N^{TOT}\langle k\rangle}{\sum\limits_{i=1}^{N_P} n_i^{eff}f\left(\frac{n_i^{eff}}{s_i}\right)}\ ,
\label{eq:Z}
\end{equation}
 where $N^{TOT}$ is the total number of individuals across the metapopulation, i.e., $N^{TOT} =\sum\limits_{j=1}^{N_P} n_j$ and $s_i$ denotes the area for node $i$. 

\begin{figure*}[t!]
    \centering
    \includegraphics[width=0.9\textwidth]{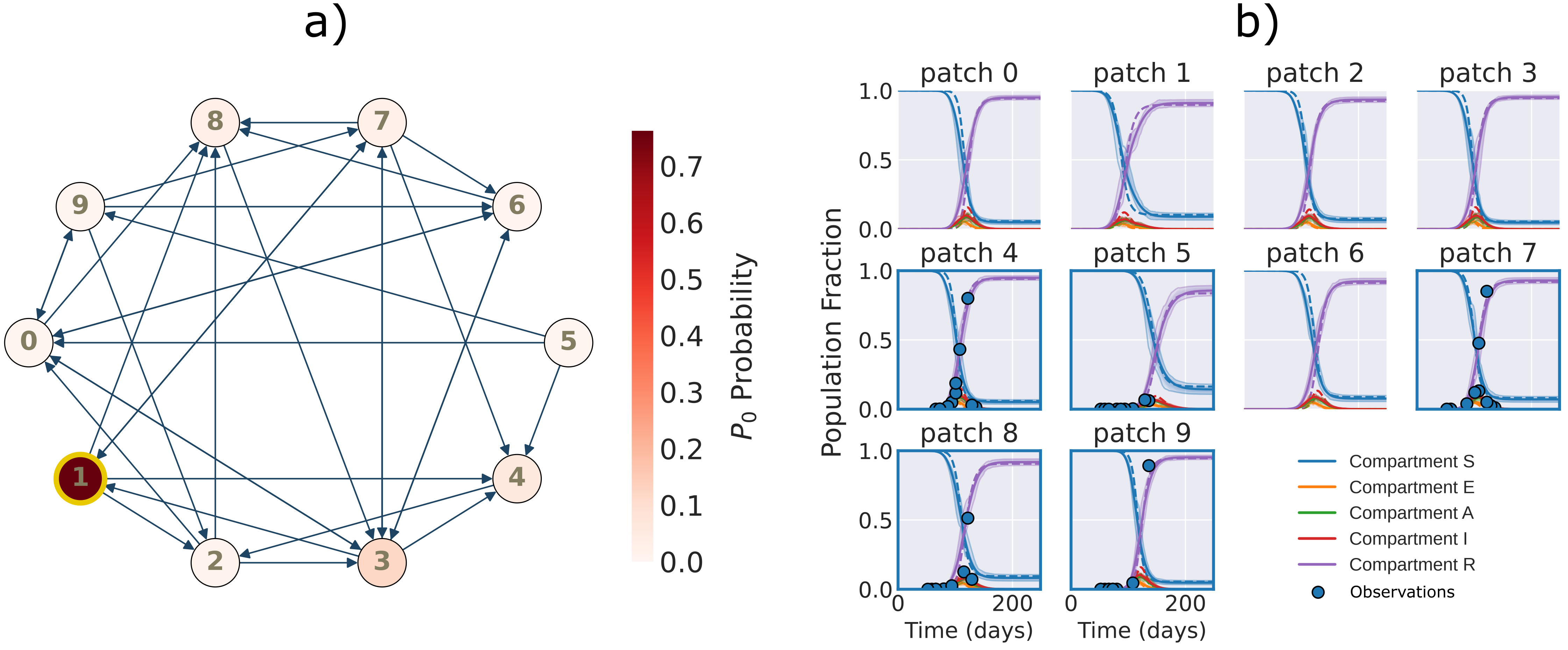}
    \caption{{\bf Predicting forward and backward dynamics in a synthetic contact network} a) A sparse synthetic metapopulation network with $N_P =10$ patches and edge-connection probabillity $\tau = 0.4$. Nodes indicate spatial regions (containing a fully-mixed population) and directed edges represent mobility flows.
    The infection is seeded in a single individual residing in Node 1 (highlighted in yellow) at time $t=0$.
    Each node is colored according to their $P_0$ probabilities (the probability of being the source of infection) as calculated by the model.
    The model predicts that node 1 is the most probable source with 76$\%$ certainty. 
    b) Dashed curves represent the \emph{ground truth} trajectories in each patch for the {\bf SEAIR} model over a period of 250-days (each time-unit $t$ is considered a single day).
    Highlighted panels and blue circles represent observations.
    Solid lines curves represent the average over the MaxEnt re-weighted ensemble of trajectories and the shaded areas represent the $\pm$33\% and $\pm$67\% quantiles.
    Model predictions match well with \emph{ground truth} trajectories ($ D_{\textrm{KL}} ^{\textrm{traj}} = 8 \times 10^{-3}$).}  \label{fig:maxent_post_network_graph}
\end{figure*}
\section{Results}
\label{sec:results}
In what it is to follow, with an initial guess on the epidemiological parameters and a set of observations, we apply our method to address two fundamental problems in epidemiology modeling: 1. Early assessment of the potential spread, 2. Identifying the origin of the outbreak.
For observations, we consider weekly averages for fraction of the population in compartments {\bf I} and {\bf R}.
We choose these two compartments, given that these are the most likely for which somewhat reliable estimates can be made from real-world data.
Nevertheless, it is well documented~\cite{lau2021evaluating} that such estimates are noisy and their fidelity varies from region to region and therefore to account for this, 
To account for the some degree of uncertainty about the data, we add multiplicative noise with a mean 1 and standard deviation 0.05 to the observations obtained from the \emph{ground truth} trajectory.
The sampling process tries to explore the trajectory space by adjusting the epidemiological parameters such as $\beta$, $\beta'$, $\epsilon$, $\eta$ and $\mu$ from normal or truncated normal distributions,while varying the infection seed across different spatial patches, as well as accounting for a small variance in the mobility flows.
Finally, Maxent re-weights the ensemble trajectories, maximizing entropy subject to the observations and determining the most probable state of the network.
We consider a Laplace distribution prior (Eq.~\ref{eq:Laplace_prior}) with standard deviation of 1 to allow some disagreement between the MaxEnt fit and the observations.
The MaxEnt implementation is done using Adam optimizer\cite{kingma2014adam} with starting learning rate of $10^{-2}$ and reduced learning rate on plateau callback (factor of 0.9, patience of 10 and minimum learning rate of $10^{-4}$) for 1000 epochs.
To assess the model's performance we compare the predictions against a  \emph{ground truth} trajectory derived from known pre-selected parameters.
Knowledge of the ground-truth enables a proof-of-concept analysis to assess model performance under different scenarios.
The ones we consider are density of the network, temporal window of observations, the number of observations and variations in mobility flow of observations with respect to the infection seeded origin.
As performance metrics we consider:
\begin{itemize}
  \item \textbf{Forward dynamics}: To compare the predicted trajectory against the known \emph{ground truth} trajectory we measure the KL-divergence, defined as   
  \begin{equation}
    \label{eq_kl_loss}
     D_{\textrm{KL}} ^{\textrm{traj}} = \frac{1}{T N_P} \sum\limits_{t=0}^T\sum\limits_{i=1}^{N_P}\sum\limits_{m} \rho^m_i(t)\log\left(\frac{\rho^m_i(t)}{\hat{\rho}^m_i(t)}\right).   
    \end{equation}
  \noindent Here $T$ is the total time in the epidemic trajectory and $m$ is the label for the compartments.
  The term $\hat{\rho}^m_i(t)$ is the model's prediction for the probability of an individual associated with patch $i$ to belong to a compartment $m$ at time $t$ and $\rho^m_i(t)$ is the corresponding value for the \emph{ground truth} trajectory. 
  
  \item \textbf{Backward dynamics}: The accuracy of the model in making the correct prediction with respect to the ground-truth source of infection ($P_0$).
  This can be treated as a binary multi-class classification problem, where the correct prediction of the true origin node is regarded as the \emph{true positive (TP)} class and every other prediction falls into the \emph{false positive (FP)} class.
  Given this, the accuracy ($\alpha$) is defined as 
  \begin{equation}
    \alpha = \frac{TP}{TP+FP}.
  \label{eq:acc}
  \end{equation} 
    The posterior probabilities $P_0$ for nodes are obtained by summing over the MaxEnt posterior weights for each node seeded as the infection source---compartment {\bf E}---in the sampled trajectories ensemble at $t=0$, and the largest value among the set corresponds to $P_0$ probability.
    To assess performance, we use the top-k posterior probabilities $P_0$, and the frequency of \emph{true positive} predictions as our metric.
     For instance, for $k=5$, the model's prediction for $P_0$ is classified as a \emph{true positive} if the infection-source is among the top five values of $P_0$ probabilities and a \emph{false positive} otherwise.
\end{itemize}
We employ our method on two systems: a synthetic metapopulation network, and the mobility network of New York state at the resolution of counties.

\subsection{Synthetic Contact Networks}
The 10-node metapopulation ($N_P = 10$) is represented as a directed graph in Fig.~\ref{fig:maxent_post_network_graph}a, where each node (patch) in the network represents a town or city in the metapopulation and the directed edges account for mobility flows between them.
The nodes are connected at random with a connection probability $\tau= 0.4$, such that on average each node is connected to four other patches (considering both in- and out-flows).
The area of each node, the population, and entries of the mobility matrix are sampled from normal distributions with parameters listed in Table~\ref{Tab:ref_param}
.
\begin{table*}[ht]
\centering
\caption{Distributions of input parameters for the synthetic contact network \emph{ground truth} trajectory.\label{Tab:ref_param}}
\begin{tabular}{|c|c|c|c|c|c|}
\hline
\bf{Parameter}&\bf{Distribution}&\bf{Mean}&\bf{Std}&\bf{Min}&\bf{Max} \\
\hline
Area&Normal&$2 \times 10^{3}$&$10^{3}$&300&--\\
\hline
Populations&Normal&$5 \times 10^{5}$&$3 \times 10^{5}$&--&--\\
\hline
$T_{ij} (i= j)$&Normal&$10^{5}$&$3 \times 10^{3}$&0&--\\
\hline
$T_{ij} (i\neq j)$&Normal&$10^{2}$&$5 \times 10^{1}$& 0&--\\
\hline
\end{tabular}
\end{table*}%

The infection is initially seeded in patch 1 (node with the yellow edge in Fig.~\ref{fig:maxent_post_network_graph}a) with a single individual exposed to the disease at $t=0$.
The parameters for this \emph{ground truth} trajectory are chosen to be: $\beta'=0.025$, $\beta=0.05$, $\epsilon=0.6$, $\eta=\frac{1}{1.2}$ and $\mu=\frac{1}{7}$.
In Fig \ref{fig:maxent_post_network_graph}b we show as dashed curves the trajectory of the \emph{ground truth} {\bf SEAIR} model for all 10 nodes for a period of $T=250$ days.
We define a distribution of the parameters and explore the trajectory space.
These distributions and their Kernel density estimation plots can be found in Table S1 and Fig. S1 in supporting material.
For all 8,192 sampled trajectories, we assume a uniform probability of infection, and randomly choose a patch, and an individual in that patch as the infection seed (see Fig. S2 in supporting material).
As observations we consider a total of 50 data points (weekly-averages) from the {\bf I} and {\bf R} compartments within an observation window of $(50,140)$ days.
The highlighted panels and blue circles in Fig.~\ref{fig:maxent_post_network_graph}b mark the five randomly chosen patches and the observations, respectively. 

We use the MaxEnt framework, to re-weight the ensemble of trajectories to agree best with the observed data-points and obtain the  $P_0$ probability by summing all the weights for each exposed node in the sampled trajectories at $t=0$. 
The re-weighted average over the sampled trajectories are shown as solid curves in Fig.~\ref{fig:maxent_post_network_graph}b, and the shaded area marks the $\pm$33\% and $\pm$67\% quantiles.
The calculated $D_{\textrm{KL}} ^{\textrm{traj}}$ of $8\times 10^{-3}$ indicates close agreement between model predictions and the \emph{ground truth} trajectory.
In Fig.~\ref{fig:maxent_post_network_graph}a we also show nodes colored by their value of $P_0$ probability, indicating that the algorithm predicts node 1 (the true-origin of infection) as the most probable source with a certainty of 76\%.

\subsubsection{Effect of network density}
Next we check the accuracy of the model as a function of the density of connections between nodes.
We tune the connection probability in the range $0.25 \leq \tau \leq 1$ to sample the spectrum between a sparse and fully-connected network.
We redo our simulations over 8,000 different networks in this range, and for each trajectory choose a random node from which to seed the infection.
\begin{figure*}[tbh!]
    \centering
    \includegraphics[width=.9\textwidth]{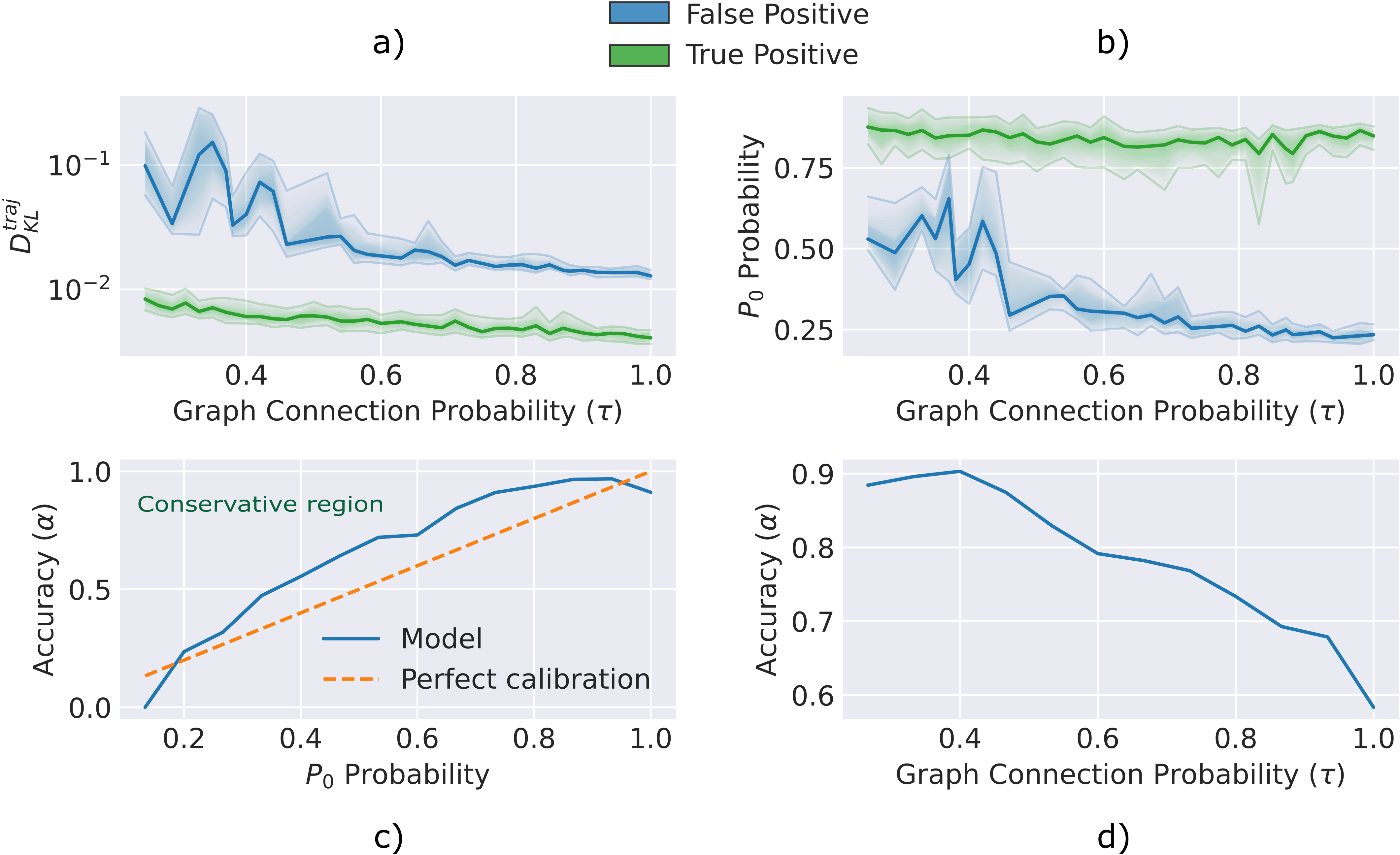}
    \caption{{\bf Effect of network density on model performance.} Epidemic-evolution on 8,000 synthetic metapopulation networks with connection probabilities in the range $0.25 \leq \tau \leq 1$. All other parameters kept the same. a) Performance of forward-dynamics, $D_{\textrm{KL}} ^{\textrm{traj}}$ as a function of $\tau$. Solid lines indicate the mode and the shaded areas mark shows the 30$\%$ confidence interval. Green marks the region where the model indicates the true-seed as the most probable infection source, and blue otherwise. b) The  $P_0$ probability as a function of $\tau$, lines and shaded regions the same as in a). c) Reliability diagram for the model, where the dashed line represents a perfectly calibrated model where accuracy $\alpha$ changes linearly with certainty. The model's predictions falls into the green shaded conservative region, suggesting that it is more accurate than it believes. d) Accuracy $\alpha$  as a function of the connection probability $\tau$, indicating a performance drop as one moves from sparse to dense graphs (given the same number of observations).}
    \label{fig:shift_p}
\end{figure*}
All other relevant parameters are kept the same. 
In Fig.\ref{fig:shift_p}a we plot $D_{\textrm{KL}} ^{\textrm{traj}}$ as a function of $\tau$, where the solid lines indicate the mode over 200 samples for a given $\tau$, and the shaded areas mark the 30$\%$ confidence interval.
The region marked in green corresponds to the \emph{True positives} (TP) where the algorithm correctly identifies the true infection-seed as the most probable source, whereas the region marked in blue corresponds to \emph{False Positives} (FP) when the true source was not identified as the most probable.
Here we use a $k=1$ acceptance criteria,  
a rather stringent condition, as even when the true source is identified as the second most probable, it is still marked FP.
The low values of $D_{\textrm{KL}} ^{\textrm{traj}}$ indicates that irrespective of the correct identification of the infection-seed, the predicted and \emph{ground truth} trajectories match well, independent of network density. 
Note that this is true for the chosen observations obtained in the $(50,140)$ day temporal window and will be further discussed later.

Additionally, we find high values of $P_0$ for TP, that is (mostly) independent of the connection probability $\tau$, while for FP, we find low values of $P_0$ that get progressively worse with increasing $\tau$ (Fig.~\ref{fig:shift_p}b).
The model's calibration is assessed in the reliability diagram shown in Fig.~\ref{fig:shift_p}c, where we plot the accuracy $\alpha$ as a function of $P_0$.
The case of a perfectly calibrated model, where $\alpha$ changes linearly with certainty is shown as the orange dashed line.
The figure indicates that the model is more accurate than it believes, in a conservative manner.
Finally, in Fig.~\ref{fig:shift_p}d  we plot $\alpha$ as a function of $\tau$ finding that the model's performance degrades in high-density networks, which is to be expected given that dense networks have more complexity in their mobility flows. Nevertheless, at worst, the model shows $\approx 60\%$ accuracy in a fully-connected graph.
Indeed, for a wide-range of connection probabilities (corresponding to realistic settings) we find an accuracy in the range of $80-90\%$.

\subsubsection{Effect of temporal window of observations}
\begin{figure*}[t!]
   \centering
    \includegraphics[width=\textwidth]{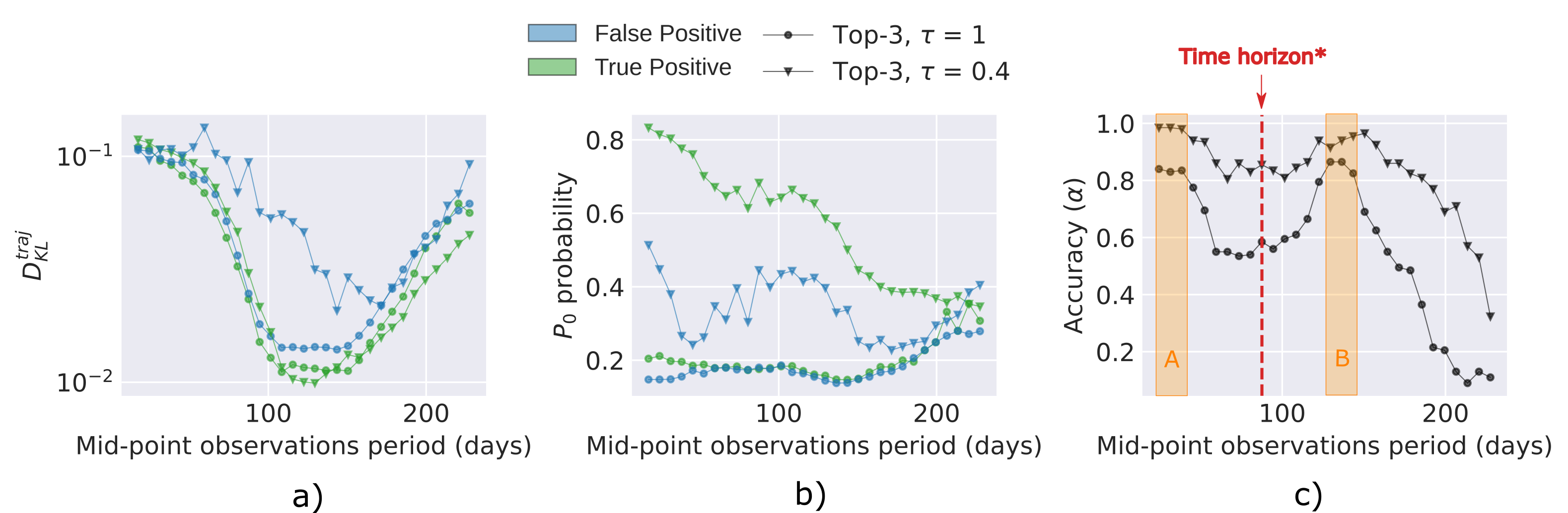}
    \caption{\label{fig:compare_shift_obs_top_3} {\bf Effect of observations temporal window on model performance.} Epidemic-evolution on 8,000 synthetic metapopulation sparse and dense networks, with $\tau$ of 0.4 and 1, respectively. Green shows true positive predictions and blue accounts for false positives in sparse (triangle) and dense (circle) networks, given a top-3 acceptance criteria. Each point in panels a and b represent the mode over 200 samples at the corresponding mid-point observation period. a) Forward dynamics predictions assessment using $D_{\textrm{KL}} ^{\textrm{traj}}$ between the MaxEnt re-weighted trajectory and \emph{ground truth}. b) Mode values for the $P_0$  probabilities. c) Accuracy vs mid-point observation period.  Model’s accuracy drops as observations are obtained from time values beyond early stages of the outbreak (stage A) but increases again at more advanced time periods (stage B) and beyond the time horizon ($t_{hor}$), where $t_{hor}$ (adapted from \cite{shah2020finding}), is a reported fundamental limit beyond which no algorithm can detect the true origin of infection.}
\end{figure*}

Next we evaluate the model's performance as a function of the temporal window in which observations are made.
Current understanding of epidemic dynamics, suggests that contact-tracing is effective only in the initial stages of the outbreak, and any information on the infection source is lost at later times.
Indeed, in~\cite{shah2020finding} an approximation to this temporal horizon, $t_{hor}$, was derived for the {\bf SIR} model. Adapting the formulation to the {\bf SEAIR} model, leads to an expression of the form:
\begin{equation}
t_{hor} = \lambda_{max}^{-1}  \log \left(\frac{N^{TOT}}{c_{max}}\right) ,
\label{eq:thor}
\end{equation}
where $\lambda_{max}$ corresponds to the leading eigenvalue of the linearized system of ODEs governing the evolution of the dynamics and $c_{max}$ a constant needed to fix the infectious seeds at the beginning of the outbreak (see Appendix~\ref{app:timehorizon} for a complete derivation). 
We consider a sparse ($\tau = 0.4$) and dense ($\tau = 1$) network and check for the presence of such a temporal horizon, by shifting the 5-week observation period within the range $T=250$, collecting 50 data points (5 points from each of compartments {\bf I} and {\bf R} for 5 random nodes).
As a robustness check, we exclude the true-infection source from our observed samples. 
In Figs.~\ref{fig:compare_shift_obs_top_3}a,b, we plot $D_{\textrm{KL}} ^{\textrm{traj}}$ and $P_0$ as a function of the mid-point of observations for each 5-week window (200 sample runs in each bin), where curves indicate the mode and shapes refer to dense (circles) and sparse (inverted triangles) networks.
Curves are split into TP (green) and FP (blue). In the figure, we show the $k=3$ acceptance criteria, and in Fig. S3 in the SI we show the case for a $k=1$ acceptance criteria.
In Fig.~\ref{fig:compare_shift_obs_top_3}c, we plot the accuracy $\alpha$ as a function of the mid-point of observations.
As expected, the figure indicates high accuracy at the early stages of the outbreak (marked Region A), and decreases as the epidemic progresses. Considering the set of parameters $(\beta,\beta^\prime,\mu,\langle k\rangle,\epsilon,\eta)=(0.05,0.025,\frac{1}{7},10,0.6,\frac{1}{1.2})$ and a seed composed of a single exposed individual at the beginning of the outbreak, we obtain $c_{max}=0.372$ and $t_{hor}=90.9$ days marked as a red vertical dashed line. 

Surprisingly, as one crosses $t_{hor}$ a non-monotonic trend is observed and a new peak in the accuracy is observed at later times ($ t \approx 150$) in both sparse and dense networks, marked as Region B.  
To the best of our knowledge, this peak in accuracy at advanced stages of the epidemic evolution, where information can be recovered on the infection source, has not been reported before.
Indeed, this region also corresponds to the lowest values of $D_{\textrm{KL}} ^{\textrm{traj}
}$ indicating the closest match to the \emph{ground truth} trajectory, and thus an optimal window in which to simultaneously infer the most accurate information in forward- and backward-dynamics (panels a and b in Fig.~\ref{fig:compare_shift_obs_top_3}, respectively).
A possible explanation for this phenomenon is that it corresponds to region with the highest gradients in epidemic curves (Fig.~\ref{fig:maxent_post_network_graph}b), whereas the low-gradients of the trajectories at other values of $t$ provides the model with insufficient information to perform a reliable inference.

\begin{figure*}[t!]
    \centering
    \includegraphics[width=\textwidth]{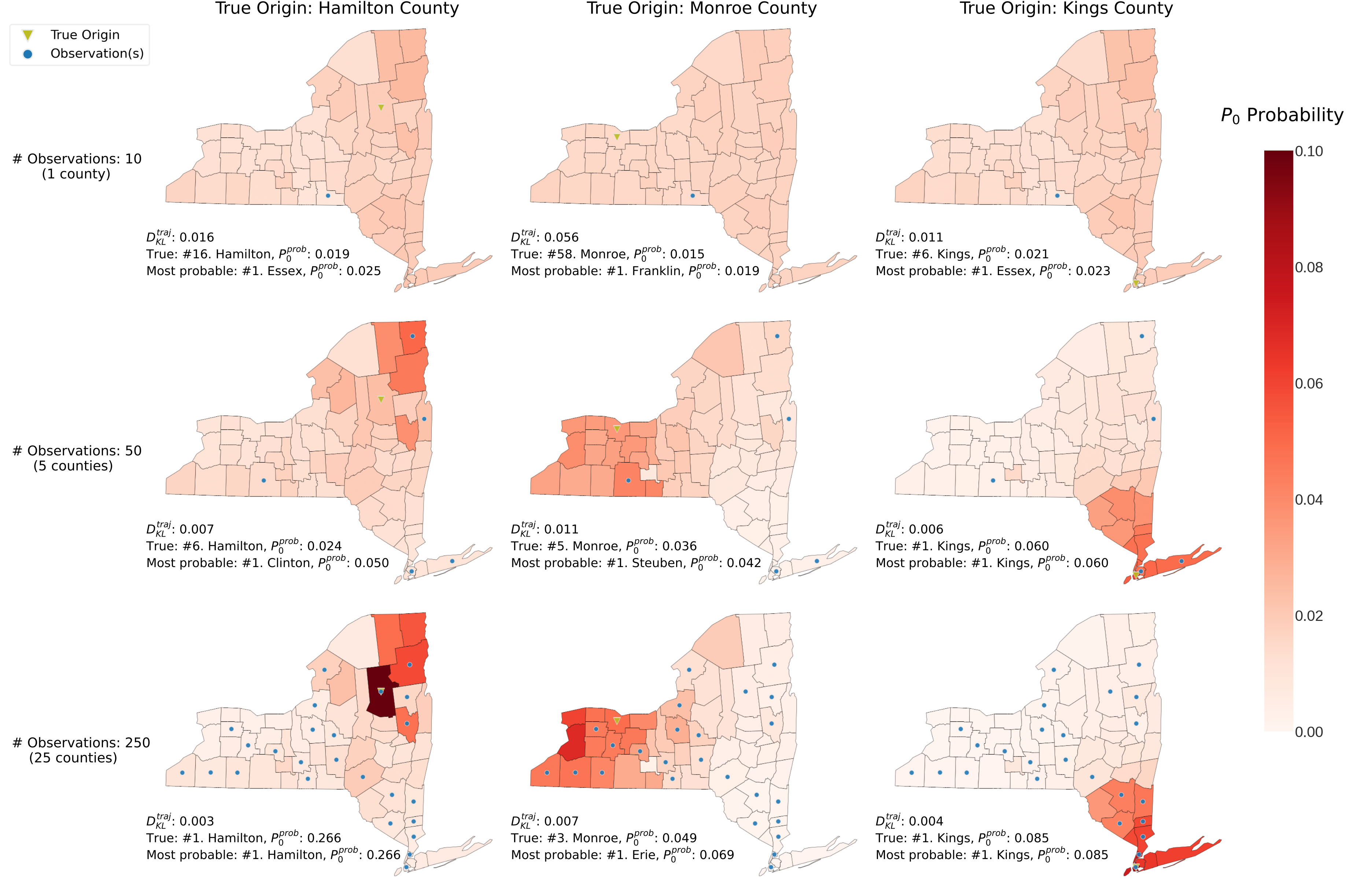}
    \caption{\label{fig:NY_network} {\bf Effect of number of observations on model performance.}
    Epidemic-evolution in the real mobility network of New York state. Yellow downward triangle represents the true origin of infection and blue circles show the counties observed.
    The $P_0$ predictions are ranked based on their probabilities.
    The most probable and true origin ranks, as well as the $D_{\textrm{KL}} ^{\textrm{traj}}$ values are shown for each simulation.
    For example, in the simulation with Kings county as the true origin with 10 observations, model's most probable prediction for the infection source is Essex county with $P_0^{prob}$ of 0.023, while the true origin is ranked 6$^{th}$ with $P_0^{prob}$ of 0.021.
    Starting from the first row to bottom, with the increase of the number of observations, model is able to infer the true origin of $P_0$ among the top-5 most probable predictions and obtain a well fit to the \emph{ground truth} trajectory, balancing both future and backward dynamics predictions. }
\end{figure*}

\subsection{Mobility Network of New York State}
In this section, we apply our formalism to characterize the spread of infectious diseases across a real metapopulation, the network of commuters across New York state at the spatial resolution of counties, of which there are 62.
The mobility flows between counties, as well as their respective areas and populations are obtained from the United States LODES commuting database~\cite{census_data}.
Our focus here is on assessing the performance of the method in detecting the spatial location of the infection-seed given more complex and realistic mobility patterns.
We first generate the \emph{ground truth} trajectory according to the following epidemic parameters: $\beta'=0.029$, $\beta=0.052$, $\epsilon=0.586$, $\eta=\frac{1}{2.493}$, $\langle k\rangle=10$ and $\mu=\frac{1}{1.49}$, and then collect observations corresponding to weekly averages of populations in compartments {\bf I} and {\bf R}.
Observations are collected from specific counties and are drawn from the $(60,140)$ day temporal window.


\subsubsection{Effect of the number of observations}
Given that the number of observations is directly linked to epidemic-surveillance efforts, we first check the performance of our model as a function of the number of counties from which data is collected.
Specifically, we test the accuracy of identifying the correct spatial origin of the infection-seed as we increase the number of counties observed. 
We choose three counties with different population densities in which to seed the infection: Hamilton ($2.74$ per square mile), Monroe ($1.14 \times 10^3$ per square mile), and Kings ($3.72 \times 10^4$ per square mile).
We collect 10 samples from each county (randomly chosen) and vary the number of counties observed from 1, 5 and 25. 
We do not necessarily exclude the seed counties from our randomly chosen observations.

In Fig.~\ref{fig:NY_network} we plot the counties colored according to their values of the posterior probability $P_0$.
The top row represents observations from a single county, the middle row from 5 counties and the bottom row 25 counties. 
The true-origin is marked as a downward yellow triangle, and the observations by blue circles.
The three columns correspond to the different infection seeds.
In each case, we show $D_{\textrm{KL}} ^{\textrm{traj}}$, $P_0$ for the true-origin and how the model ranks it as a likely source of infection, as well as the models prediction for the top-ranked county in terms of the posterior probability $P_0$.
For all three infection-sources, observations from a single county yields poor results for $D_{\textrm{KL}} ^{\textrm{traj}}$, and the model ranks the true-origin quite low as a probable source (16 for Hamilton, 58 for Monroe and 6 for Kings).
Sampling from 5 counties results in a considerable increase in performance for the first two counties (6 for Hamilton, 5 for Monroe) while for Kings the model correctly identifies it at the most likely origin.
We also note about an order of magnitude decrease in $D_{\textrm{KL}} ^{\textrm{traj}}$ for all three counties indicating good agreement with the forward dynamics. 
Finally, sampling from 25 counties results in the best performance where in addition to Kings, the model correctly identifies Hamilton as a true infection source, while for the case of Monroe the model ranks it as the third most likely origin. 
We see further improvements in matching the forward dynamics with further decreases in $D_{\textrm{KL}} ^{\textrm{traj}}$ (about two orders of magnitude as compared to observing as single county). 
As an illustrative example we show the full trajectory-set for Monroe county true-origin with 250 observations in Fig. S4 in the supporting material.

We note the difference in accuracy of the model when assessing Hamilton and Monroe counties.
Hamilton despite being a much more sparsely populated area than Monroe, was correctly identified as the true source, whereas Monroe was ranked third.
The reason for this discrepancy is that Hamilton was also included in the sample of 25 counties as an input to the model, whereas Monroe was excluded from its observation set.
The likelihood of the model to correctly guess the true source increases greatly when the source itself is included as an observation, a feature also seen in our synthetic metapopulation networks. 
On the other hand, the ability of the model to identify Monroe as the third most likely source is notable given that no information on Monroe was available to the model.
Indeed, Erie county, adjacent to Monroe was marked as the most likely source of infection.
Kings county is an outlier compared to the other two, in that already with a single observed county the model marked it amongst the upper $10\%$ of posterior probabilities $P_0$.
Certainly there are more people in Kings (it has the highest population density by far among the three counties) but also it is coterminous with Brooklyn, and a popular destination for residents of other counties.
Therefore there is a higher likelihood of mixing of populations from different parts of the state. 

\subsubsection{Dependence of accuracy on effective proximity}
\begin{figure}[t!]
    \centering
    \includegraphics[width=\columnwidth]{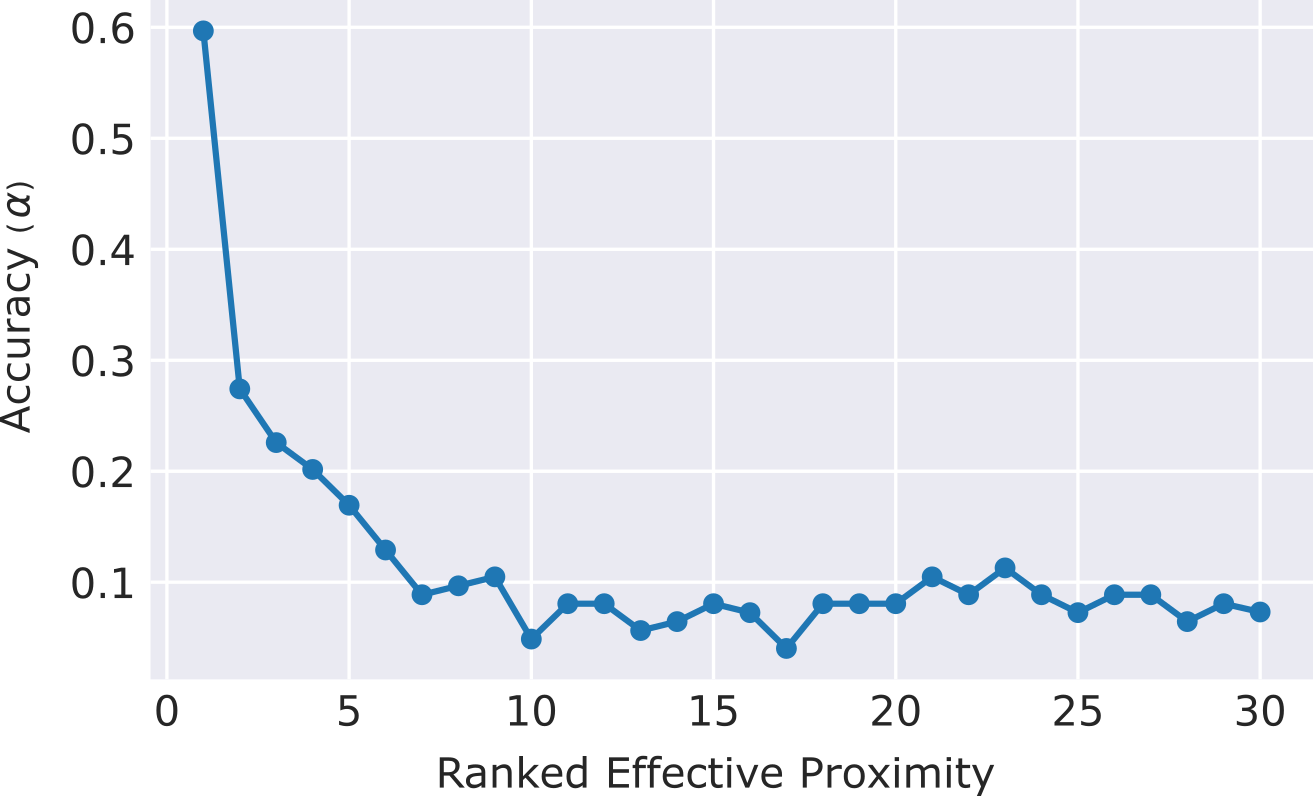}
    \caption{\label{fig:NY_mobility_ranking} {\bf Effect of mobility-strength between counties on accuracy in the mobility network of New York state}. The accuracy of the model as a function of the \emph{effective proximity} ${\bf \phi}$ (Eq.~\eqref{eq:mobility_flow_strength}. Each point represents the average over 180 runs of the model with a randomly selected true origin county and 10 sampled trajectories from a single county. The horizontal axis corresponds to the ranking of the values of $\phi$. Rank-1 corresponds to the case when the observations are made from the source county.}
\end{figure}

Given the latter observation, we next check whether the strength of mobility flows (both in and out) between counties plays a role in the model's  accuracy. 
Two locations are strongly connected if there are many people traveling between them, and therefore we define an \emph{effective proximity} matrix ${\bf \phi}$ with elements given by
\begin{equation}
    \phi_{ij} = \frac{1}{ R_{ij} + R_{ji}},
    \label{eq:mobility_flow_strength}
\end{equation}
where ${\bf R}$ is the OD matrix, and we take into account both in- and out-flows.
In this setting counties that are strongly connected by mobility flows have low values of $\phi_{ij}$ and are therefore more proximal in mobility space.
We next seed the infection in location $i$ and sample from a single county $j$ (including the source), ranked in increasing order according to their value of $\phi_{ij}$ with the rank of $i$ corresponding to 1.
We then generate 8,000 trajectories with a randomly sampled true origin, and plot the accuracy $\alpha$ as a function of effective proximity to the origin county in Fig.~\ref{fig:NY_mobility_ranking}. 
Each point in the figure corresponds to the average over 180 realizations. 
We clearly see a monotonically decreasing trend;  sampling from counties further away from the origin-county leads to a sharp decline in accuracy saturating at around the 7th furthest county. 
The trend is expected given that locations further away from the source in mobility space, experience delays in arrivals of infectious cases.
This lag results in the observation of degenerate epidemic trajectories, thus making the inference less accurate.

\section{Conclusions}
\label{sec:discuss}
This paper has provided, to the best of the authors' knowledge, the first systematic study of both backward and forward dynamics inference on contagion process in contact networks.
We have applied the statistical mechanics principle of maximum entropy to the conventional {\bf SEAIR} epidemiology models to re-weight disease trajectories and obtain the best fit to a set of observations, while making reliable predictions on the true source of the outbreak.
The novelty of this work lies within working well under the sparse-data regime and highly uncertain initial parameter priors, making our method highly suitable for studying disease dynamics. 
Finally, the method proposed here is independent of the underlying compartmental model.
While we presented our work in the context of epidemics, the approach is easily generalizable to similar classes of spreading processes.
For example, a single computer virus can infect millions of other computers through the Internet.
An isolated failure in an electrical power grid network can result a city-wide blackout.
Misinformation or a baleful rumor can spread through social networks and cause terror and inconvenience.
In all these scenarios, the contagion process \cite{stroock2007multidimensional, pastor2015epidemic} could identify the source of the risk on the network and quarantine its harmful effects \cite{centola2007complex, baronchelli2018emergence, wang2013modeling, mishra2013mathematical}.

\section*{Acknowledgements}
We thank the Center for Integrated Research Computing (CIRC) at University of Rochester for providing computational resources and technical support. Funding for this research was provided by National Science Foundation under Grant No. 2029095. D.S.-P. acknowledges financial support from Spanish Ministerio de Ciencia e Innovación (projects FIS2017-87519-P and PID2020-113582GB-I00), from the Departamento de Industria e Innovación del Gobierno de Aragón y Fondo Social Europeo (FENOL group E-19), and from Fundación Ibercaja and Universidad de Zaragoza (grant No.\ 224220).

\section*{Code Availability Statement}
The  MaxEnt  implementation  is  publicly  available  on  Github (\href{https://github.com/ur-whitelab/maxent}{\color{blue}{\underline{https://github.com/ur-whitelab/maxent}}})  as  a  python  package  called {\bf maxent} and it can be applied to any simulator. The {\bf SEAIR} model used in this work is publicly available as python package called {\bf py0} on Github (\href{https://github.com/ur-whitelab/py0}{\color{blue}{\underline{https://github.com/ur-whitelab/py0}}}).
\appendix
\section{Derivation for Time Horizon}
\label{app:timehorizon}
For the sake of comparison, we now compute the $t_{hor}$ value for our compartmental model, according to the rationale followed in~\cite{shah2020finding}. Mathematically, the authors define the time horizon as the time at which the number of infectious individuals, whose evolution is assumed to follow the early stage dynamics of the outbreak, scales to the entire population. To simplify the analysis, we make a mean-field approach and neglect the contact heterogeneities existing across the different patches of the metapopulation. At this limit, the dynamics is completely characterized by the fraction of the population in each compartment $m$ at each time step $t$, denoted in the following by $\rho^m(t)$. Specifically:

\begin{eqnarray}
\rho^S(t+1) &=& (1-\Pi(t))\rho^S(t) \label{eq:modelnS}\\
\rho^E(t+1) &=& (1-\eta)\rho^E(t) + \rho^S(t)\Pi(t)\label{eq:modelnE}\\
\rho^A(t+1) &=& (1-\epsilon)\eta\rho^E(t)+(1-\mu)\rho^A(t)\ ,\\
\rho^I(t+1) &=& \epsilon\eta\rho^E(t) + (1-\mu)\rho^I(t)\ ,\\
\rho^R(t+1) &=& \rho^R(t) +\mu \left[\rho^I(t)+\rho^A (t)\right]\label{eq:modelnD}\ ,
\end{eqnarray}
with 
\begin{equation}
\Pi (t) = 1-(1-\beta)^{\langle k\rangle\rho^I(t)}(1-\beta^\prime)^{\langle k\rangle \rho^A(t)}\ ,
\end{equation}
At the early stages of the outbreak, the number of affected individuals is negligible compared with the size of the population. Therefore, we can assume that $\rho^m \ll 1$, with $m=\left\lbrace E,A,I,R \right\rbrace$. This turns the latter expression into:
\begin{equation}
\Pi (t) \simeq \langle k\rangle (\beta\rho^I (t) + \beta^\prime \rho^A (t) )\ ,
\end{equation}
where we have considered that $\beta,\beta^\prime \ll 1$ as well. Introducing the latter expression into Eq.~(\ref{eq:modelnE}) and neglecting ${\cal O} (\rho^2)$ terms lead to
\begin{equation}
\rho^E(t+1) = (1-\eta)\rho^E(t) + \beta \langle k \rangle \rho^I (t) + \beta^\prime \langle k\rangle \rho^A (t)\ .\label{eq:modellE}
\end{equation}
For the sake of simplicity, it is convenient at this point to rewrite the equations in terms of the occupation of each compartment $m$, denoted by $m(t)$. In particular, restricting ourselves to the infectious or potentially infectious individuals, we have that
\begin{eqnarray}
\dot{E} &=& -\eta E + \beta^\prime \langle k \rangle A  +\beta \langle k \rangle I\ ,\label{eq:modellpopE}\\
\dot{A} &=& (1-\epsilon)\eta E-\mu)A\ ,\\
\dot{I}&=& \epsilon\eta E -\mu) I\ ,
\end{eqnarray}
where we have defined $\dot{m} = m(t+1) - m(t)$. Consequently, the evolution of the system is given by:
\begin{equation}
    \vec{x}(t)=\sum\limits_{i=1}^3 c_i \vec{v}_i e^{\lambda_i t}\ ,
\end{equation}
being $\lambda_i$ and $\vec{v}_i$ each of the eigenvalues and their associated eigenvectors respectively and $c_i$ the integration constants needed to fix the initial conditions to run the dynamics.
Albeit the latter expression constitutes the exact evolution of the system, the long-term dynamics is completely determined by the largest eigenvalue $\lambda_{max}$ and its associated eigenvector $\vec{v}_{max}$. Therefore, we can assume that:
\begin{equation}
  \vec{x}(t)\approx c_{max} \vec{v}_{max} e^{\lambda_{max} t}\ , 
  \label{eq:thor}
\end{equation}
with
\begin{equation}
    \lambda_{max}=\frac{\sqrt{(\eta-\mu)^{2}+4 \langle k\rangle \eta\left((1-\epsilon) \beta^{\prime}+\epsilon \beta\right)}-(\eta+\mu)}{2}\ .
\end{equation}
Without loss of generality, we set the component of the eigenvector associated with the symptomatic infectious compartment to $v_{max}^I=1$.
Finally, equating the number of symptomatic infectious individuals to the population size, we derive the time horizon $t_{hor}$ which reads as:
\begin{equation}
t_{hor} = \lambda_{max}^{-1}\log\left(\frac{N^{TOT}}{c_{max}}\right)\ .
\end{equation}
\bibliography{bibliography}

\begin{thebibliography}{97}%
\makeatletter
\providecommand \@ifxundefined [1]{%
 \@ifx{#1\undefined}
}%
\providecommand \@ifnum [1]{%
 \ifnum #1\expandafter \@firstoftwo
 \else \expandafter \@secondoftwo
 \fi
}%
\providecommand \@ifx [1]{%
 \ifx #1\expandafter \@firstoftwo
 \else \expandafter \@secondoftwo
 \fi
}%
\providecommand \natexlab [1]{#1}%
\providecommand \enquote  [1]{``#1''}%
\providecommand \bibnamefont  [1]{#1}%
\providecommand \bibfnamefont [1]{#1}%
\providecommand \citenamefont [1]{#1}%
\providecommand \href@noop [0]{\@secondoftwo}%
\providecommand \href [0]{\begingroup \@sanitize@url \@href}%
\providecommand \@href[1]{\@@startlink{#1}\@@href}%
\providecommand \@@href[1]{\endgroup#1\@@endlink}%
\providecommand \@sanitize@url [0]{\catcode `\\12\catcode `\$12\catcode
  `\&12\catcode `\#12\catcode `\^12\catcode `\_12\catcode `\%12\relax}%
\providecommand \@@startlink[1]{}%
\providecommand \@@endlink[0]{}%
\providecommand \url  [0]{\begingroup\@sanitize@url \@url }%
\providecommand \@url [1]{\endgroup\@href {#1}{\urlprefix }}%
\providecommand \urlprefix  [0]{URL }%
\providecommand \Eprint [0]{\href }%
\providecommand \doibase [0]{http://dx.doi.org/}%
\providecommand \selectlanguage [0]{\@gobble}%
\providecommand \bibinfo  [0]{\@secondoftwo}%
\providecommand \bibfield  [0]{\@secondoftwo}%
\providecommand \translation [1]{[#1]}%
\providecommand \BibitemOpen [0]{}%
\providecommand \bibitemStop [0]{}%
\providecommand \bibitemNoStop [0]{.\EOS\space}%
\providecommand \EOS [0]{\spacefactor3000\relax}%
\providecommand \BibitemShut  [1]{\csname bibitem#1\endcsname}%
\let\auto@bib@innerbib\@empty
\bibitem [{\citenamefont {Estrada}(2020)}]{Estrada2020}%
  \BibitemOpen
  \bibfield  {author} {\bibinfo {author} {\bibfnamefont {E.}~\bibnamefont
  {Estrada}},\ }\href@noop {} {\bibfield  {journal} {\bibinfo  {journal}
  {Physics Reports}\ }\textbf {\bibinfo {volume} {869}},\ \bibinfo {pages} {1}
  (\bibinfo {year} {2020})}\BibitemShut {NoStop}%
\bibitem [{\citenamefont {Organization}\ \emph {et~al.}(2020)\citenamefont
  {Organization} \emph {et~al.}}]{world2020coronavirus}%
  \BibitemOpen
  \bibfield  {author} {\bibinfo {author} {\bibfnamefont {W.~H.}\ \bibnamefont
  {Organization}} \emph {et~al.},\ }\href@noop {} {\emph {\bibinfo {title}
  {Coronavirus disease 2019 (COVID-19): situation report, 72}}},\ \bibinfo
  {type} {Tech. Rep.}\ (\bibinfo {year} {2020})\BibitemShut {NoStop}%
\bibitem [{\citenamefont {Hazarie}\ \emph {et~al.}(2021)\citenamefont
  {Hazarie}, \citenamefont {Soriano-Pa{\~n}os}, \citenamefont {Arenas},
  \citenamefont {G{\'o}mez-Garde{\~n}es},\ and\ \citenamefont
  {Ghoshal}}]{Hazarie_2021}%
  \BibitemOpen
  \bibfield  {author} {\bibinfo {author} {\bibfnamefont {S.}~\bibnamefont
  {Hazarie}}, \bibinfo {author} {\bibfnamefont {D.}~\bibnamefont
  {Soriano-Pa{\~n}os}}, \bibinfo {author} {\bibfnamefont {A.}~\bibnamefont
  {Arenas}}, \bibinfo {author} {\bibfnamefont {J.}~\bibnamefont
  {G{\'o}mez-Garde{\~n}es}}, \ and\ \bibinfo {author} {\bibfnamefont
  {G.}~\bibnamefont {Ghoshal}},\ }\href {\doibase 10.1038/s42005-021-00679-0}
  {\bibfield  {journal} {\bibinfo  {journal} {Communications Physics}\ }\textbf
  {\bibinfo {volume} {4}},\ \bibinfo {pages} {191} (\bibinfo {year}
  {2021})}\BibitemShut {NoStop}%
\bibitem [{\citenamefont {Candido}\ \emph {et~al.}(2020)\citenamefont
  {Candido}, \citenamefont {Watts}, \citenamefont {Abade}, \citenamefont
  {Kraemer}, \citenamefont {Pybus}, \citenamefont {Croda}, \citenamefont
  {De~Oliveira}, \citenamefont {Khan}, \citenamefont {Sabino},\ and\
  \citenamefont {Faria}}]{candido2020routes}%
  \BibitemOpen
  \bibfield  {author} {\bibinfo {author} {\bibfnamefont {D.~D.~S.}\
  \bibnamefont {Candido}}, \bibinfo {author} {\bibfnamefont {A.}~\bibnamefont
  {Watts}}, \bibinfo {author} {\bibfnamefont {L.}~\bibnamefont {Abade}},
  \bibinfo {author} {\bibfnamefont {M.~U.}\ \bibnamefont {Kraemer}}, \bibinfo
  {author} {\bibfnamefont {O.~G.}\ \bibnamefont {Pybus}}, \bibinfo {author}
  {\bibfnamefont {J.}~\bibnamefont {Croda}}, \bibinfo {author} {\bibfnamefont
  {W.}~\bibnamefont {De~Oliveira}}, \bibinfo {author} {\bibfnamefont
  {K.}~\bibnamefont {Khan}}, \bibinfo {author} {\bibfnamefont {E.~C.}\
  \bibnamefont {Sabino}}, \ and\ \bibinfo {author} {\bibfnamefont {N.~R.}\
  \bibnamefont {Faria}},\ }\href@noop {} {\bibfield  {journal} {\bibinfo
  {journal} {Journal of Travel Medicine}\ }\textbf {\bibinfo {volume} {27}},\
  \bibinfo {pages} {taaa042} (\bibinfo {year} {2020})}\BibitemShut {NoStop}%
\bibitem [{\citenamefont {Chinazzi}\ \emph {et~al.}(2020)\citenamefont
  {Chinazzi}, \citenamefont {Davis}, \citenamefont {Ajelli}, \citenamefont
  {Gioannini}, \citenamefont {Litvinova}, \citenamefont {Merler}, \citenamefont
  {y~Piontti}, \citenamefont {Mu}, \citenamefont {Rossi}, \citenamefont {Sun}
  \emph {et~al.}}]{chinazzi2020effect}%
  \BibitemOpen
  \bibfield  {author} {\bibinfo {author} {\bibfnamefont {M.}~\bibnamefont
  {Chinazzi}}, \bibinfo {author} {\bibfnamefont {J.~T.}\ \bibnamefont {Davis}},
  \bibinfo {author} {\bibfnamefont {M.}~\bibnamefont {Ajelli}}, \bibinfo
  {author} {\bibfnamefont {C.}~\bibnamefont {Gioannini}}, \bibinfo {author}
  {\bibfnamefont {M.}~\bibnamefont {Litvinova}}, \bibinfo {author}
  {\bibfnamefont {S.}~\bibnamefont {Merler}}, \bibinfo {author} {\bibfnamefont
  {A.~P.}\ \bibnamefont {y~Piontti}}, \bibinfo {author} {\bibfnamefont
  {K.}~\bibnamefont {Mu}}, \bibinfo {author} {\bibfnamefont {L.}~\bibnamefont
  {Rossi}}, \bibinfo {author} {\bibfnamefont {K.}~\bibnamefont {Sun}},  \emph
  {et~al.},\ }\href@noop {} {\bibfield  {journal} {\bibinfo  {journal}
  {Science}\ }\textbf {\bibinfo {volume} {368}},\ \bibinfo {pages} {395}
  (\bibinfo {year} {2020})}\BibitemShut {NoStop}%
\bibitem [{\citenamefont {Gilbert}\ \emph {et~al.}(2020)\citenamefont
  {Gilbert}, \citenamefont {Pullano}, \citenamefont {Pinotti}, \citenamefont
  {Valdano}, \citenamefont {Poletto}, \citenamefont {Bo{\"e}lle}, \citenamefont
  {d'Ortenzio}, \citenamefont {Yazdanpanah}, \citenamefont {Eholie},
  \citenamefont {Altmann} \emph {et~al.}}]{gilbert2020preparedness}%
  \BibitemOpen
  \bibfield  {author} {\bibinfo {author} {\bibfnamefont {M.}~\bibnamefont
  {Gilbert}}, \bibinfo {author} {\bibfnamefont {G.}~\bibnamefont {Pullano}},
  \bibinfo {author} {\bibfnamefont {F.}~\bibnamefont {Pinotti}}, \bibinfo
  {author} {\bibfnamefont {E.}~\bibnamefont {Valdano}}, \bibinfo {author}
  {\bibfnamefont {C.}~\bibnamefont {Poletto}}, \bibinfo {author} {\bibfnamefont
  {P.-Y.}\ \bibnamefont {Bo{\"e}lle}}, \bibinfo {author} {\bibfnamefont
  {E.}~\bibnamefont {d'Ortenzio}}, \bibinfo {author} {\bibfnamefont
  {Y.}~\bibnamefont {Yazdanpanah}}, \bibinfo {author} {\bibfnamefont {S.~P.}\
  \bibnamefont {Eholie}}, \bibinfo {author} {\bibfnamefont {M.}~\bibnamefont
  {Altmann}},  \emph {et~al.},\ }\href@noop {} {\bibfield  {journal} {\bibinfo
  {journal} {The Lancet}\ }\textbf {\bibinfo {volume} {395}},\ \bibinfo {pages}
  {871} (\bibinfo {year} {2020})}\BibitemShut {NoStop}%
\bibitem [{\citenamefont {Abedi}\ \emph {et~al.}(2021)\citenamefont {Abedi},
  \citenamefont {Olulana}, \citenamefont {Avula}, \citenamefont {Chaudhary},
  \citenamefont {Khan}, \citenamefont {Shahjouei}, \citenamefont {Li},\ and\
  \citenamefont {Zand}}]{abedi2021racial}%
  \BibitemOpen
  \bibfield  {author} {\bibinfo {author} {\bibfnamefont {V.}~\bibnamefont
  {Abedi}}, \bibinfo {author} {\bibfnamefont {O.}~\bibnamefont {Olulana}},
  \bibinfo {author} {\bibfnamefont {V.}~\bibnamefont {Avula}}, \bibinfo
  {author} {\bibfnamefont {D.}~\bibnamefont {Chaudhary}}, \bibinfo {author}
  {\bibfnamefont {A.}~\bibnamefont {Khan}}, \bibinfo {author} {\bibfnamefont
  {S.}~\bibnamefont {Shahjouei}}, \bibinfo {author} {\bibfnamefont
  {J.}~\bibnamefont {Li}}, \ and\ \bibinfo {author} {\bibfnamefont
  {R.}~\bibnamefont {Zand}},\ }\href@noop {} {\bibfield  {journal} {\bibinfo
  {journal} {Journal of racial and ethnic health disparities}\ }\textbf
  {\bibinfo {volume} {8}},\ \bibinfo {pages} {732} (\bibinfo {year}
  {2021})}\BibitemShut {NoStop}%
\bibitem [{\citenamefont {Ahmed}\ \emph {et~al.}(2020)\citenamefont {Ahmed},
  \citenamefont {Ahmed}, \citenamefont {Pissarides},\ and\ \citenamefont
  {Stiglitz}}]{ahmed2020inequality}%
  \BibitemOpen
  \bibfield  {author} {\bibinfo {author} {\bibfnamefont {F.}~\bibnamefont
  {Ahmed}}, \bibinfo {author} {\bibfnamefont {N.}~\bibnamefont {Ahmed}},
  \bibinfo {author} {\bibfnamefont {C.}~\bibnamefont {Pissarides}}, \ and\
  \bibinfo {author} {\bibfnamefont {J.}~\bibnamefont {Stiglitz}},\ }\href@noop
  {} {\bibfield  {journal} {\bibinfo  {journal} {The Lancet Public Health}\
  }\textbf {\bibinfo {volume} {5}},\ \bibinfo {pages} {e240} (\bibinfo {year}
  {2020})}\BibitemShut {NoStop}%
\bibitem [{\citenamefont {Barbosa}\ \emph {et~al.}(2021)\citenamefont
  {Barbosa}, \citenamefont {Hazarie}, \citenamefont {Dickinson}, \citenamefont
  {Bassolas}, \citenamefont {Frank}, \citenamefont {Kautz}, \citenamefont
  {Sadilek}, \citenamefont {Ramasco},\ and\ \citenamefont
  {Ghoshal}}]{Barbosa_2021}%
  \BibitemOpen
  \bibfield  {author} {\bibinfo {author} {\bibfnamefont {H.}~\bibnamefont
  {Barbosa}}, \bibinfo {author} {\bibfnamefont {S.}~\bibnamefont {Hazarie}},
  \bibinfo {author} {\bibfnamefont {B.}~\bibnamefont {Dickinson}}, \bibinfo
  {author} {\bibfnamefont {A.}~\bibnamefont {Bassolas}}, \bibinfo {author}
  {\bibfnamefont {A.}~\bibnamefont {Frank}}, \bibinfo {author} {\bibfnamefont
  {H.}~\bibnamefont {Kautz}}, \bibinfo {author} {\bibfnamefont
  {A.}~\bibnamefont {Sadilek}}, \bibinfo {author} {\bibfnamefont
  {J.}~\bibnamefont {Ramasco}}, \ and\ \bibinfo {author} {\bibfnamefont
  {G.}~\bibnamefont {Ghoshal}},\ }\href {\doibase 10.1038/s41598-021-87407-4}
  {\bibfield  {journal} {\bibinfo  {journal} {Scientific Reports}\ }\textbf
  {\bibinfo {volume} {11}},\ \bibinfo {pages} {8616} (\bibinfo {year}
  {2021})}\BibitemShut {NoStop}%
\bibitem [{\citenamefont {Berry}(2008)}]{berry2008urbanization}%
  \BibitemOpen
  \bibfield  {author} {\bibinfo {author} {\bibfnamefont {B.~J.}\ \bibnamefont
  {Berry}},\ }in\ \href@noop {} {\emph {\bibinfo {booktitle} {Urban ecology}}}\
  (\bibinfo  {publisher} {Springer},\ \bibinfo {year} {2008})\ pp.\ \bibinfo
  {pages} {25--48}\BibitemShut {NoStop}%
\bibitem [{\citenamefont {Bertinelli}\ and\ \citenamefont
  {Black}(2004)}]{bertinelli2004urbanization}%
  \BibitemOpen
  \bibfield  {author} {\bibinfo {author} {\bibfnamefont {L.}~\bibnamefont
  {Bertinelli}}\ and\ \bibinfo {author} {\bibfnamefont {D.}~\bibnamefont
  {Black}},\ }\href@noop {} {\bibfield  {journal} {\bibinfo  {journal} {Journal
  of Urban Economics}\ }\textbf {\bibinfo {volume} {56}},\ \bibinfo {pages}
  {80} (\bibinfo {year} {2004})}\BibitemShut {NoStop}%
\bibitem [{\citenamefont {Kraemer}\ \emph {et~al.}(2020)\citenamefont
  {Kraemer}, \citenamefont {Yang}, \citenamefont {Gutierrez}, \citenamefont
  {Wu}, \citenamefont {Klein}, \citenamefont {Pigott}, \citenamefont
  {Du~Plessis}, \citenamefont {Faria}, \citenamefont {Li}, \citenamefont
  {Hanage} \emph {et~al.}}]{kraemer2020}%
  \BibitemOpen
  \bibfield  {author} {\bibinfo {author} {\bibfnamefont {M.~U.}\ \bibnamefont
  {Kraemer}}, \bibinfo {author} {\bibfnamefont {C.-H.}\ \bibnamefont {Yang}},
  \bibinfo {author} {\bibfnamefont {B.}~\bibnamefont {Gutierrez}}, \bibinfo
  {author} {\bibfnamefont {C.-H.}\ \bibnamefont {Wu}}, \bibinfo {author}
  {\bibfnamefont {B.}~\bibnamefont {Klein}}, \bibinfo {author} {\bibfnamefont
  {D.~M.}\ \bibnamefont {Pigott}}, \bibinfo {author} {\bibfnamefont
  {L.}~\bibnamefont {Du~Plessis}}, \bibinfo {author} {\bibfnamefont {N.~R.}\
  \bibnamefont {Faria}}, \bibinfo {author} {\bibfnamefont {R.}~\bibnamefont
  {Li}}, \bibinfo {author} {\bibfnamefont {W.~P.}\ \bibnamefont {Hanage}},
  \emph {et~al.},\ }\href@noop {} {\bibfield  {journal} {\bibinfo  {journal}
  {Science}\ }\textbf {\bibinfo {volume} {368}},\ \bibinfo {pages} {493}
  (\bibinfo {year} {2020})}\BibitemShut {NoStop}%
\bibitem [{\citenamefont {Maier}\ and\ \citenamefont
  {Brockmann}(2020)}]{maier2020}%
  \BibitemOpen
  \bibfield  {author} {\bibinfo {author} {\bibfnamefont {B.~F.}\ \bibnamefont
  {Maier}}\ and\ \bibinfo {author} {\bibfnamefont {D.}~\bibnamefont
  {Brockmann}},\ }\href@noop {} {\bibfield  {journal} {\bibinfo  {journal}
  {Science}\ }\textbf {\bibinfo {volume} {368}},\ \bibinfo {pages} {742}
  (\bibinfo {year} {2020})}\BibitemShut {NoStop}%
\bibitem [{\citenamefont {Gatto}\ \emph {et~al.}(2020)\citenamefont {Gatto},
  \citenamefont {Bertuzzo}, \citenamefont {Mari}, \citenamefont {Miccoli},
  \citenamefont {Carraro}, \citenamefont {Casagrandi},\ and\ \citenamefont
  {Rinaldo}}]{gatto2020}%
  \BibitemOpen
  \bibfield  {author} {\bibinfo {author} {\bibfnamefont {M.}~\bibnamefont
  {Gatto}}, \bibinfo {author} {\bibfnamefont {E.}~\bibnamefont {Bertuzzo}},
  \bibinfo {author} {\bibfnamefont {L.}~\bibnamefont {Mari}}, \bibinfo {author}
  {\bibfnamefont {S.}~\bibnamefont {Miccoli}}, \bibinfo {author} {\bibfnamefont
  {L.}~\bibnamefont {Carraro}}, \bibinfo {author} {\bibfnamefont
  {R.}~\bibnamefont {Casagrandi}}, \ and\ \bibinfo {author} {\bibfnamefont
  {A.}~\bibnamefont {Rinaldo}},\ }\href {\doibase 10.1073/pnas.2004978117}
  {\bibfield  {journal} {\bibinfo  {journal} {Procs. Nat. Acad. Sci. U.S.A.}\
  }\textbf {\bibinfo {volume} {117}},\ \bibinfo {pages} {10484} (\bibinfo
  {year} {2020})}\BibitemShut {NoStop}%
\bibitem [{\citenamefont {Aleta}\ \emph {et~al.}(2020)\citenamefont {Aleta},
  \citenamefont {Martin-Corral}, \citenamefont {Pastore~y Piontti},
  \citenamefont {Ajelli}, \citenamefont {Litvinova}, \citenamefont {Chinazzi},
  \citenamefont {Dean}, \citenamefont {Halloran}, \citenamefont {Longini},
  \citenamefont {Merler}, \citenamefont {Pentland}, \citenamefont {Vespignani},
  \citenamefont {Moro},\ and\ \citenamefont {Moreno}}]{aleta2020}%
  \BibitemOpen
  \bibfield  {author} {\bibinfo {author} {\bibfnamefont {A.}~\bibnamefont
  {Aleta}}, \bibinfo {author} {\bibfnamefont {D.}~\bibnamefont
  {Martin-Corral}}, \bibinfo {author} {\bibfnamefont {A.}~\bibnamefont
  {Pastore~y Piontti}}, \bibinfo {author} {\bibfnamefont {M.}~\bibnamefont
  {Ajelli}}, \bibinfo {author} {\bibfnamefont {M.}~\bibnamefont {Litvinova}},
  \bibinfo {author} {\bibfnamefont {M.}~\bibnamefont {Chinazzi}}, \bibinfo
  {author} {\bibfnamefont {N.~E.}\ \bibnamefont {Dean}}, \bibinfo {author}
  {\bibfnamefont {M.~E.}\ \bibnamefont {Halloran}}, \bibinfo {author}
  {\bibfnamefont {I.~M.}\ \bibnamefont {Longini}}, \bibinfo {author}
  {\bibfnamefont {S.}~\bibnamefont {Merler}}, \bibinfo {author} {\bibfnamefont
  {A.}~\bibnamefont {Pentland}}, \bibinfo {author} {\bibfnamefont
  {A.}~\bibnamefont {Vespignani}}, \bibinfo {author} {\bibfnamefont
  {E.}~\bibnamefont {Moro}}, \ and\ \bibinfo {author} {\bibfnamefont
  {Y.}~\bibnamefont {Moreno}},\ }\href@noop {} {\bibfield  {journal} {\bibinfo
  {journal} {medRxiv}\ ,\ \bibinfo {pages} {2020.05.06.20092841}} (\bibinfo
  {year} {2020})}\BibitemShut {NoStop}%
\bibitem [{\citenamefont {Dehning}\ \emph {et~al.}(2020)\citenamefont
  {Dehning}, \citenamefont {Zierenberg}, \citenamefont {Spitzner},
  \citenamefont {Wibral}, \citenamefont {Neto}, \citenamefont {Wilczek},\ and\
  \citenamefont {Priesemann}}]{dehning2020}%
  \BibitemOpen
  \bibfield  {author} {\bibinfo {author} {\bibfnamefont {J.}~\bibnamefont
  {Dehning}}, \bibinfo {author} {\bibfnamefont {J.}~\bibnamefont {Zierenberg}},
  \bibinfo {author} {\bibfnamefont {F.~P.}\ \bibnamefont {Spitzner}}, \bibinfo
  {author} {\bibfnamefont {M.}~\bibnamefont {Wibral}}, \bibinfo {author}
  {\bibfnamefont {J.~P.}\ \bibnamefont {Neto}}, \bibinfo {author}
  {\bibfnamefont {M.}~\bibnamefont {Wilczek}}, \ and\ \bibinfo {author}
  {\bibfnamefont {V.}~\bibnamefont {Priesemann}},\ }\href {\doibase
  10.1126/science.abb9789} {\bibfield  {journal} {\bibinfo  {journal}
  {Science}\ }\textbf {\bibinfo {volume} {369}},\ \bibinfo {pages} {160}
  (\bibinfo {year} {2020})}\BibitemShut {NoStop}%
\bibitem [{\citenamefont {Anderson}\ and\ \citenamefont
  {May}(1992)}]{anderson1992infectious}%
  \BibitemOpen
  \bibfield  {author} {\bibinfo {author} {\bibfnamefont {R.~M.}\ \bibnamefont
  {Anderson}}\ and\ \bibinfo {author} {\bibfnamefont {R.~M.}\ \bibnamefont
  {May}},\ }\href@noop {} {\emph {\bibinfo {title} {Infectious diseases of
  humans: dynamics and control}}}\ (\bibinfo  {publisher} {Oxford university
  press},\ \bibinfo {year} {1992})\BibitemShut {NoStop}%
\bibitem [{\citenamefont {Keeling}\ and\ \citenamefont
  {Rohani}(2011)}]{keeling2011modeling}%
  \BibitemOpen
  \bibfield  {author} {\bibinfo {author} {\bibfnamefont {M.~J.}\ \bibnamefont
  {Keeling}}\ and\ \bibinfo {author} {\bibfnamefont {P.}~\bibnamefont
  {Rohani}},\ }\href@noop {} {\emph {\bibinfo {title} {Modeling infectious
  diseases in humans and animals}}}\ (\bibinfo  {publisher} {Princeton
  University Press},\ \bibinfo {year} {2011})\BibitemShut {NoStop}%
\bibitem [{\citenamefont {Vynnycky}\ and\ \citenamefont
  {White}(2010)}]{vynnycky2010introduction}%
  \BibitemOpen
  \bibfield  {author} {\bibinfo {author} {\bibfnamefont {E.}~\bibnamefont
  {Vynnycky}}\ and\ \bibinfo {author} {\bibfnamefont {R.}~\bibnamefont
  {White}},\ }\href@noop {} {\emph {\bibinfo {title} {An introduction to
  infectious disease modelling}}}\ (\bibinfo  {publisher} {OUP oxford},\
  \bibinfo {year} {2010})\BibitemShut {NoStop}%
\bibitem [{\citenamefont {Meehan}\ \emph {et~al.}(2020)\citenamefont {Meehan},
  \citenamefont {Rojas}, \citenamefont {Adekunle}, \citenamefont {Adegboye},
  \citenamefont {Caldwell}, \citenamefont {Turek}, \citenamefont {Williams},
  \citenamefont {Trauer},\ and\ \citenamefont {McBryde}}]{meehan2020modelling}%
  \BibitemOpen
  \bibfield  {author} {\bibinfo {author} {\bibfnamefont {M.~T.}\ \bibnamefont
  {Meehan}}, \bibinfo {author} {\bibfnamefont {D.~P.}\ \bibnamefont {Rojas}},
  \bibinfo {author} {\bibfnamefont {A.~I.}\ \bibnamefont {Adekunle}}, \bibinfo
  {author} {\bibfnamefont {O.~A.}\ \bibnamefont {Adegboye}}, \bibinfo {author}
  {\bibfnamefont {J.~M.}\ \bibnamefont {Caldwell}}, \bibinfo {author}
  {\bibfnamefont {E.}~\bibnamefont {Turek}}, \bibinfo {author} {\bibfnamefont
  {B.}~\bibnamefont {Williams}}, \bibinfo {author} {\bibfnamefont {J.~M.}\
  \bibnamefont {Trauer}}, \ and\ \bibinfo {author} {\bibfnamefont {E.~S.}\
  \bibnamefont {McBryde}},\ }\href@noop {} {\bibfield  {journal} {\bibinfo
  {journal} {Paediatric respiratory reviews}\ } (\bibinfo {year}
  {2020})}\BibitemShut {NoStop}%
\bibitem [{\citenamefont {Roosa}\ and\ \citenamefont
  {Chowell}(2019)}]{roosa2019assessing}%
  \BibitemOpen
  \bibfield  {author} {\bibinfo {author} {\bibfnamefont {K.}~\bibnamefont
  {Roosa}}\ and\ \bibinfo {author} {\bibfnamefont {G.}~\bibnamefont
  {Chowell}},\ }\href@noop {} {\bibfield  {journal} {\bibinfo  {journal}
  {Theoretical Biology and Medical Modelling}\ }\textbf {\bibinfo {volume}
  {16}},\ \bibinfo {pages} {1} (\bibinfo {year} {2019})}\BibitemShut {NoStop}%
\bibitem [{\citenamefont {Castro}\ \emph {et~al.}(2020)\citenamefont {Castro},
  \citenamefont {Ares}, \citenamefont {Cuesta},\ and\ \citenamefont
  {Manrubia}}]{castro2020turning}%
  \BibitemOpen
  \bibfield  {author} {\bibinfo {author} {\bibfnamefont {M.}~\bibnamefont
  {Castro}}, \bibinfo {author} {\bibfnamefont {S.}~\bibnamefont {Ares}},
  \bibinfo {author} {\bibfnamefont {J.~A.}\ \bibnamefont {Cuesta}}, \ and\
  \bibinfo {author} {\bibfnamefont {S.}~\bibnamefont {Manrubia}},\ }\href@noop
  {} {\bibfield  {journal} {\bibinfo  {journal} {Proceedings of the National
  Academy of Sciences}\ }\textbf {\bibinfo {volume} {117}},\ \bibinfo {pages}
  {26190} (\bibinfo {year} {2020})}\BibitemShut {NoStop}%
\bibitem [{\citenamefont {Wilke}\ and\ \citenamefont
  {Bergstrom}(2020)}]{wilke2020predicting}%
  \BibitemOpen
  \bibfield  {author} {\bibinfo {author} {\bibfnamefont {C.~O.}\ \bibnamefont
  {Wilke}}\ and\ \bibinfo {author} {\bibfnamefont {C.~T.}\ \bibnamefont
  {Bergstrom}},\ }\href@noop {} {\bibfield  {journal} {\bibinfo  {journal}
  {Proceedings of the National Academy of Sciences}\ }\textbf {\bibinfo
  {volume} {117}},\ \bibinfo {pages} {28549} (\bibinfo {year}
  {2020})}\BibitemShut {NoStop}%
\bibitem [{\citenamefont {Jewell}\ \emph {et~al.}(2020)\citenamefont {Jewell},
  \citenamefont {Lewnard},\ and\ \citenamefont
  {Jewell}}]{jewell2020predictive}%
  \BibitemOpen
  \bibfield  {author} {\bibinfo {author} {\bibfnamefont {N.~P.}\ \bibnamefont
  {Jewell}}, \bibinfo {author} {\bibfnamefont {J.~A.}\ \bibnamefont {Lewnard}},
  \ and\ \bibinfo {author} {\bibfnamefont {B.~L.}\ \bibnamefont {Jewell}},\
  }\href@noop {} {\bibfield  {journal} {\bibinfo  {journal} {Jama}\ }\textbf
  {\bibinfo {volume} {323}},\ \bibinfo {pages} {1893} (\bibinfo {year}
  {2020})}\BibitemShut {NoStop}%
\bibitem [{\citenamefont {Pandey}\ \emph {et~al.}(2013)\citenamefont {Pandey},
  \citenamefont {Mubayi},\ and\ \citenamefont {Medlock}}]{Pandey2013DengueSIR}%
  \BibitemOpen
  \bibfield  {author} {\bibinfo {author} {\bibfnamefont {A.}~\bibnamefont
  {Pandey}}, \bibinfo {author} {\bibfnamefont {A.}~\bibnamefont {Mubayi}}, \
  and\ \bibinfo {author} {\bibfnamefont {J.}~\bibnamefont {Medlock}},\ }\href
  {\doibase 10.1016/j.mbs.2013.10.007} {\bibfield  {journal} {\bibinfo
  {journal} {Mathematical Biosciences}\ }\textbf {\bibinfo {volume} {246}},\
  \bibinfo {pages} {252} (\bibinfo {year} {2013})}\BibitemShut {NoStop}%
\bibitem [{\citenamefont {Lavielle}\ \emph {et~al.}(2011)\citenamefont
  {Lavielle}, \citenamefont {Samson}, \citenamefont {{Karina Fermin}},\ and\
  \citenamefont {Mentr{\'{e}}}}]{Lavielle2011HIVSIR}%
  \BibitemOpen
  \bibfield  {author} {\bibinfo {author} {\bibfnamefont {M.}~\bibnamefont
  {Lavielle}}, \bibinfo {author} {\bibfnamefont {A.}~\bibnamefont {Samson}},
  \bibinfo {author} {\bibfnamefont {A.}~\bibnamefont {{Karina Fermin}}}, \ and\
  \bibinfo {author} {\bibfnamefont {F.}~\bibnamefont {Mentr{\'{e}}}},\ }\href
  {\doibase 10.1111/j.1541-0420.2010.01422.x} {\bibfield  {journal} {\bibinfo
  {journal} {Biometrics}\ }\textbf {\bibinfo {volume} {67}},\ \bibinfo {pages}
  {250} (\bibinfo {year} {2011})}\BibitemShut {NoStop}%
\bibitem [{\citenamefont {New}\ \emph {et~al.}(2009)\citenamefont {New},
  \citenamefont {Matthiopoulos}, \citenamefont {Redpath},\ and\ \citenamefont
  {Buckland}}]{New2009SIRMCMC}%
  \BibitemOpen
  \bibfield  {author} {\bibinfo {author} {\bibfnamefont {L.~F.}\ \bibnamefont
  {New}}, \bibinfo {author} {\bibfnamefont {J.}~\bibnamefont {Matthiopoulos}},
  \bibinfo {author} {\bibfnamefont {S.}~\bibnamefont {Redpath}}, \ and\
  \bibinfo {author} {\bibfnamefont {S.~T.}\ \bibnamefont {Buckland}},\ }\href
  {\doibase 10.1086/603625} {\bibfield  {journal} {\bibinfo  {journal}
  {American Naturalist}\ }\textbf {\bibinfo {volume} {174}},\ \bibinfo {pages}
  {399} (\bibinfo {year} {2009})}\BibitemShut {NoStop}%
\bibitem [{\citenamefont {Morton}\ and\ \citenamefont
  {Finkenstadt}(2005)}]{Morton2005SIRMCMC}%
  \BibitemOpen
  \bibfield  {author} {\bibinfo {author} {\bibfnamefont {A.}~\bibnamefont
  {Morton}}\ and\ \bibinfo {author} {\bibfnamefont {B.~F.}\ \bibnamefont
  {Finkenstadt}},\ }\href {\doibase 10.1111/j.1467-9876.2005.05366.x}
  {\bibfield  {journal} {\bibinfo  {journal} {Journal of the Royal Statistical
  Society: Series C (Applied Statistics)}\ }\textbf {\bibinfo {volume} {54}},\
  \bibinfo {pages} {575} (\bibinfo {year} {2005})}\BibitemShut {NoStop}%
\bibitem [{\citenamefont {Cauchemez}\ and\ \citenamefont
  {Ferguson}(2008)}]{Cauchemez2008SIRMCMC}%
  \BibitemOpen
  \bibfield  {author} {\bibinfo {author} {\bibfnamefont {S.}~\bibnamefont
  {Cauchemez}}\ and\ \bibinfo {author} {\bibfnamefont {N.~M.}\ \bibnamefont
  {Ferguson}},\ }\href {\doibase 10.1098/rsif.2007.1292} {\bibfield  {journal}
  {\bibinfo  {journal} {Journal of The Royal Society Interface}\ }\textbf
  {\bibinfo {volume} {5}},\ \bibinfo {pages} {885} (\bibinfo {year}
  {2008})}\BibitemShut {NoStop}%
\bibitem [{\citenamefont {Talawar}\ and\ \citenamefont
  {Aundhakar}(2016)}]{Talawar2016SIRMCMC}%
  \BibitemOpen
  \bibfield  {author} {\bibinfo {author} {\bibfnamefont {A.~S.}\ \bibnamefont
  {Talawar}}\ and\ \bibinfo {author} {\bibfnamefont {U.~R.}\ \bibnamefont
  {Aundhakar}},\ }\href {http://www.ripublication.com} {\emph {\bibinfo {title}
  {Global Journal of Pure and Applied Mathematics}}},\ \bibinfo {type} {Tech.
  Rep.}\ \bibinfo {number} {2}\ (\bibinfo {year} {2016})\BibitemShut {NoStop}%
\bibitem [{\citenamefont {He}\ \emph {et~al.}(2010)\citenamefont {He},
  \citenamefont {Ionides},\ and\ \citenamefont {King}}]{He2010SIRSMC}%
  \BibitemOpen
  \bibfield  {author} {\bibinfo {author} {\bibfnamefont {D.}~\bibnamefont
  {He}}, \bibinfo {author} {\bibfnamefont {E.~L.}\ \bibnamefont {Ionides}}, \
  and\ \bibinfo {author} {\bibfnamefont {A.~A.}\ \bibnamefont {King}},\ }\href
  {\doibase 10.1098/rsif.2009.0151} {\bibfield  {journal} {\bibinfo  {journal}
  {Journal of The Royal Society Interface}\ }\textbf {\bibinfo {volume} {7}},\
  \bibinfo {pages} {271} (\bibinfo {year} {2010})}\BibitemShut {NoStop}%
\bibitem [{\citenamefont {King}\ \emph {et~al.}(2008)\citenamefont {King},
  \citenamefont {Ionides}, \citenamefont {Pascual},\ and\ \citenamefont
  {Bouma}}]{King2008SIRSMC}%
  \BibitemOpen
  \bibfield  {author} {\bibinfo {author} {\bibfnamefont {A.~A.}\ \bibnamefont
  {King}}, \bibinfo {author} {\bibfnamefont {E.~L.}\ \bibnamefont {Ionides}},
  \bibinfo {author} {\bibfnamefont {M.}~\bibnamefont {Pascual}}, \ and\
  \bibinfo {author} {\bibfnamefont {M.~J.}\ \bibnamefont {Bouma}},\ }\href
  {\doibase 10.1038/nature07084} {\bibfield  {journal} {\bibinfo  {journal}
  {Nature}\ }\textbf {\bibinfo {volume} {454}},\ \bibinfo {pages} {877}
  (\bibinfo {year} {2008})}\BibitemShut {NoStop}%
\bibitem [{\citenamefont {Ionides}\ \emph {et~al.}(2006)\citenamefont
  {Ionides}, \citenamefont {Bret{\'{o}}},\ and\ \citenamefont
  {King}}]{Ionides2006SIRSMC}%
  \BibitemOpen
  \bibfield  {author} {\bibinfo {author} {\bibfnamefont {E.~L.}\ \bibnamefont
  {Ionides}}, \bibinfo {author} {\bibfnamefont {C.}~\bibnamefont
  {Bret{\'{o}}}}, \ and\ \bibinfo {author} {\bibfnamefont {A.~A.}\ \bibnamefont
  {King}},\ }\href {\doibase 10.1073/pnas.0603181103} {\bibfield  {journal}
  {\bibinfo  {journal} {Proceedings of the National Academy of Sciences of the
  United States of America}\ }\textbf {\bibinfo {volume} {103}},\ \bibinfo
  {pages} {18438} (\bibinfo {year} {2006})}\BibitemShut {NoStop}%
\bibitem [{\citenamefont {Bock}(1983)}]{bock1983TrajectoryMatching}%
  \BibitemOpen
  \bibfield  {author} {\bibinfo {author} {\bibfnamefont {H.}~\bibnamefont
  {Bock}},\ }\href@noop {} {\bibfield  {journal} {\bibinfo  {journal} {Recent
  advances in parameter identification techniques for ode Heidelberg, Federal
  Republic of Germany}\ } (\bibinfo {year} {1983})}\BibitemShut {NoStop}%
\bibitem [{\citenamefont {Arora}\ and\ \citenamefont
  {Biegler}(2004)}]{Arora2004TrajectoryMatching}%
  \BibitemOpen
  \bibfield  {author} {\bibinfo {author} {\bibfnamefont {N.}~\bibnamefont
  {Arora}}\ and\ \bibinfo {author} {\bibfnamefont {L.~T.}\ \bibnamefont
  {Biegler}},\ }\href {\doibase 10.1023/B:COAP.0000018879.40214.11} {\bibfield
  {journal} {\bibinfo  {journal} {Computational Optimization and Applications}\
  }\textbf {\bibinfo {volume} {28}},\ \bibinfo {pages} {51} (\bibinfo {year}
  {2004})}\BibitemShut {NoStop}%
\bibitem [{\citenamefont {Banks}\ \emph {et~al.}(1981)\citenamefont {Banks},
  \citenamefont {Burns{\$}},\ and\ \citenamefont
  {Cliff}}]{Banks1981TrajectoryMatching}%
  \BibitemOpen
  \bibfield  {author} {\bibinfo {author} {\bibfnamefont {H.~T.}\ \bibnamefont
  {Banks}}, \bibinfo {author} {\bibfnamefont {J.~A.}\ \bibnamefont
  {Burns{\$}}}, \ and\ \bibinfo {author} {\bibfnamefont {E.~M.}\ \bibnamefont
  {Cliff}},\ }\href {http://www.siam.org/journals/ojsa.php} {\emph {\bibinfo
  {title} {SIAM J. CONTROL AND OPTIMIZATION}}},\ \bibinfo {type} {Tech. Rep.}\
  \bibinfo {number} {6}\ (\bibinfo {year} {1981})\BibitemShut {NoStop}%
\bibitem [{\citenamefont {Ciupe}\ \emph {et~al.}(2006)\citenamefont {Ciupe},
  \citenamefont {Bivort}, \citenamefont {Bortz},\ and\ \citenamefont
  {Nelson}}]{Ciupe2006TrajectoryMatching}%
  \BibitemOpen
  \bibfield  {author} {\bibinfo {author} {\bibfnamefont {M.~S.}\ \bibnamefont
  {Ciupe}}, \bibinfo {author} {\bibfnamefont {B.~L.}\ \bibnamefont {Bivort}},
  \bibinfo {author} {\bibfnamefont {D.~M.}\ \bibnamefont {Bortz}}, \ and\
  \bibinfo {author} {\bibfnamefont {P.~W.}\ \bibnamefont {Nelson}},\ }\href
  {\doibase 10.1016/j.mbs.2005.12.006} {\bibfield  {journal} {\bibinfo
  {journal} {Mathematical Biosciences}\ }\textbf {\bibinfo {volume} {200}},\
  \bibinfo {pages} {1} (\bibinfo {year} {2006})}\BibitemShut {NoStop}%
\bibitem [{\citenamefont {Biegler}\ \emph {et~al.}(1986)\citenamefont
  {Biegler}, \citenamefont {Damiano},\ and\ \citenamefont
  {Blau}}]{Biegler1986TrajectoryMatching}%
  \BibitemOpen
  \bibfield  {author} {\bibinfo {author} {\bibfnamefont {L.~T.}\ \bibnamefont
  {Biegler}}, \bibinfo {author} {\bibfnamefont {J.~J.}\ \bibnamefont
  {Damiano}}, \ and\ \bibinfo {author} {\bibfnamefont {G.~E.}\ \bibnamefont
  {Blau}},\ }\href {\doibase 10.1002/aic.690320105} {\bibfield  {journal}
  {\bibinfo  {journal} {AIChE Journal}\ }\textbf {\bibinfo {volume} {32}},\
  \bibinfo {pages} {29} (\bibinfo {year} {1986})}\BibitemShut {NoStop}%
\bibitem [{\citenamefont {Yang}\ \emph {et~al.}(2013)\citenamefont {Yang},
  \citenamefont {Liu}, \citenamefont {Xu}, \citenamefont {Yang}, \citenamefont
  {Dong},\ and\ \citenamefont {Zhang}}]{Yang2013LeastSquaresSVM}%
  \BibitemOpen
  \bibfield  {author} {\bibinfo {author} {\bibfnamefont {R.}~\bibnamefont
  {Yang}}, \bibinfo {author} {\bibfnamefont {R.}~\bibnamefont {Liu}}, \bibinfo
  {author} {\bibfnamefont {K.}~\bibnamefont {Xu}}, \bibinfo {author}
  {\bibfnamefont {Y.}~\bibnamefont {Yang}}, \bibinfo {author} {\bibfnamefont
  {G.}~\bibnamefont {Dong}}, \ and\ \bibinfo {author} {\bibfnamefont
  {W.}~\bibnamefont {Zhang}},\ }\href {\doibase 10.1039/c3ay41014e} {\bibfield
  {journal} {\bibinfo  {journal} {Analytical Methods}\ }\textbf {\bibinfo
  {volume} {5}},\ \bibinfo {pages} {5949} (\bibinfo {year} {2013})}\BibitemShut
  {NoStop}%
\bibitem [{\citenamefont {Hooker}\ \emph {et~al.}(2011)\citenamefont {Hooker},
  \citenamefont {Ellner}, \citenamefont {Roditi},\ and\ \citenamefont
  {Earn}}]{Hooker2011SIRGeneralizedProfiling}%
  \BibitemOpen
  \bibfield  {author} {\bibinfo {author} {\bibfnamefont {G.}~\bibnamefont
  {Hooker}}, \bibinfo {author} {\bibfnamefont {S.~P.}\ \bibnamefont {Ellner}},
  \bibinfo {author} {\bibfnamefont {L.~D.~V.}\ \bibnamefont {Roditi}}, \ and\
  \bibinfo {author} {\bibfnamefont {D.~J.~D.}\ \bibnamefont {Earn}},\ }\href
  {\doibase 10.1098/rsif.2010.0412} {\bibfield  {journal} {\bibinfo  {journal}
  {Journal of The Royal Society Interface}\ }\textbf {\bibinfo {volume} {8}},\
  \bibinfo {pages} {961} (\bibinfo {year} {2011})}\BibitemShut {NoStop}%
\bibitem [{\citenamefont {Blum}\ and\ \citenamefont
  {Tran}(2010)}]{Blum2010SIRABC}%
  \BibitemOpen
  \bibfield  {author} {\bibinfo {author} {\bibfnamefont {M.~G.~B.}\
  \bibnamefont {Blum}}\ and\ \bibinfo {author} {\bibfnamefont {C.}~\bibnamefont
  {Tran}},\ }\href {\doibase 10.1093/biostatistics/kxq022} {\bibfield
  {journal} {\bibinfo  {journal} {Biostatistics}\ }\textbf {\bibinfo {volume}
  {11}},\ \bibinfo {pages} {644} (\bibinfo {year} {2010})}\BibitemShut
  {NoStop}%
\bibitem [{\citenamefont {Toni}\ \emph {et~al.}(2009)\citenamefont {Toni},
  \citenamefont {Welch}, \citenamefont {Strelkowa}, \citenamefont {Ipsen},\
  and\ \citenamefont {Stumpf}}]{Toni2009SIRABC}%
  \BibitemOpen
  \bibfield  {author} {\bibinfo {author} {\bibfnamefont {T.}~\bibnamefont
  {Toni}}, \bibinfo {author} {\bibfnamefont {D.}~\bibnamefont {Welch}},
  \bibinfo {author} {\bibfnamefont {N.}~\bibnamefont {Strelkowa}}, \bibinfo
  {author} {\bibfnamefont {A.}~\bibnamefont {Ipsen}}, \ and\ \bibinfo {author}
  {\bibfnamefont {M.~P.}\ \bibnamefont {Stumpf}},\ }\href {\doibase
  10.1098/rsif.2008.0172} {\bibfield  {journal} {\bibinfo  {journal} {Journal
  of The Royal Society Interface}\ }\textbf {\bibinfo {volume} {6}},\ \bibinfo
  {pages} {187} (\bibinfo {year} {2009})}\BibitemShut {NoStop}%
\bibitem [{\citenamefont {Kypraios}\ \emph {et~al.}(2017)\citenamefont
  {Kypraios}, \citenamefont {Neal},\ and\ \citenamefont
  {Prangle}}]{Kypraios2017SIRABC}%
  \BibitemOpen
  \bibfield  {author} {\bibinfo {author} {\bibfnamefont {T.}~\bibnamefont
  {Kypraios}}, \bibinfo {author} {\bibfnamefont {P.}~\bibnamefont {Neal}}, \
  and\ \bibinfo {author} {\bibfnamefont {D.}~\bibnamefont {Prangle}},\ }\href
  {\doibase 10.1016/j.mbs.2016.07.001} {\bibfield  {journal} {\bibinfo
  {journal} {Mathematical Biosciences}\ }\textbf {\bibinfo {volume} {287}},\
  \bibinfo {pages} {42} (\bibinfo {year} {2017})}\BibitemShut {NoStop}%
\bibitem [{\citenamefont {Rogalsky}(2012)}]{Rogalsky2012SIRGradientFree}%
  \BibitemOpen
  \bibfield  {author} {\bibinfo {author} {\bibfnamefont {T.}~\bibnamefont
  {Rogalsky}},\ }\href@noop {} {\emph {\bibinfo {title} {{B{\'{e}}zier Control
  Parameterization for Evolutionary Optimization in Disease Models}}}}\
  (\bibinfo {year} {2012})\BibitemShut {NoStop}%
\bibitem [{\citenamefont {Iacoviello}\ \emph {et~al.}(2008)\citenamefont
  {Iacoviello}, \citenamefont {Iacoviello},\ and\ \citenamefont
  {Liuzzi}}]{Iacoviello2008SIRGradientFree}%
  \BibitemOpen
  \bibfield  {author} {\bibinfo {author} {\bibfnamefont {D.}~\bibnamefont
  {Iacoviello}}, \bibinfo {author} {\bibfnamefont {D.}~\bibnamefont
  {Iacoviello}}, \ and\ \bibinfo {author} {\bibfnamefont {G.}~\bibnamefont
  {Liuzzi}},\ }\href {\doibase 10.2507/IJSIMM07(2)3.103} {\bibfield  {journal}
  {\bibinfo  {journal} {Article in International Journal of Simulation
  Modelling}\ } (\bibinfo {year} {2008}),\
  10.2507/IJSIMM07(2)3.103}\BibitemShut {NoStop}%
\bibitem [{\citenamefont {Streftaris}\ and\ \citenamefont
  {Gibson}(2004)}]{Streftaris2004SIRBayesianInference}%
  \BibitemOpen
  \bibfield  {author} {\bibinfo {author} {\bibfnamefont {G.}~\bibnamefont
  {Streftaris}}\ and\ \bibinfo {author} {\bibfnamefont {G.~J.}\ \bibnamefont
  {Gibson}},\ }\href {\doibase 10.1191/1471082X04st065oa} {\bibfield  {journal}
  {\bibinfo  {journal} {Statistical Modelling: An International Journal}\
  }\textbf {\bibinfo {volume} {4}},\ \bibinfo {pages} {63} (\bibinfo {year}
  {2004})}\BibitemShut {NoStop}%
\bibitem [{\citenamefont {Clancy}\ and\ \citenamefont
  {O'Neill}(2008)}]{Clancy2008SIRBayesianInference}%
  \BibitemOpen
  \bibfield  {author} {\bibinfo {author} {\bibfnamefont {D.}~\bibnamefont
  {Clancy}}\ and\ \bibinfo {author} {\bibfnamefont {P.~D.}\ \bibnamefont
  {O'Neill}},\ }\href {\doibase 10.1214/08-BA328} {\bibfield  {journal}
  {\bibinfo  {journal} {Bayesian Analysis}\ }\textbf {\bibinfo {volume} {3}},\
  \bibinfo {pages} {737} (\bibinfo {year} {2008})}\BibitemShut {NoStop}%
\bibitem [{\citenamefont {DEMIRIS}\ and\ \citenamefont
  {O'NEILL}(2005)}]{DEMIRIS2005SIRBayesianInference}%
  \BibitemOpen
  \bibfield  {author} {\bibinfo {author} {\bibfnamefont {N.}~\bibnamefont
  {DEMIRIS}}\ and\ \bibinfo {author} {\bibfnamefont {P.~D.}\ \bibnamefont
  {O'NEILL}},\ }\href {\doibase 10.1111/j.1467-9469.2005.00420.x} {\bibfield
  {journal} {\bibinfo  {journal} {Scandinavian Journal of Statistics}\ }\textbf
  {\bibinfo {volume} {32}},\ \bibinfo {pages} {265} (\bibinfo {year}
  {2005})}\BibitemShut {NoStop}%
\bibitem [{\citenamefont {Altarelli}\ \emph {et~al.}(2014)\citenamefont
  {Altarelli}, \citenamefont {Braunstein}, \citenamefont {Dall'Asta},
  \citenamefont {Lage-Castellanos},\ and\ \citenamefont
  {Zecchina}}]{Altarelli2014SIRBayesianInference}%
  \BibitemOpen
  \bibfield  {author} {\bibinfo {author} {\bibfnamefont {F.}~\bibnamefont
  {Altarelli}}, \bibinfo {author} {\bibfnamefont {A.}~\bibnamefont
  {Braunstein}}, \bibinfo {author} {\bibfnamefont {L.}~\bibnamefont
  {Dall'Asta}}, \bibinfo {author} {\bibfnamefont {A.}~\bibnamefont
  {Lage-Castellanos}}, \ and\ \bibinfo {author} {\bibfnamefont
  {R.}~\bibnamefont {Zecchina}},\ }\href {\doibase
  10.1103/PhysRevLett.112.118701} {\bibfield  {journal} {\bibinfo  {journal}
  {Physical Review Letters}\ }\textbf {\bibinfo {volume} {112}},\ \bibinfo
  {pages} {118701} (\bibinfo {year} {2014})},\ \Eprint
  {http://arxiv.org/abs/1307.6786} {arXiv:1307.6786} \BibitemShut {NoStop}%
\bibitem [{\citenamefont {{El Maroufy}}\ \emph {et~al.}(2016)\citenamefont {{El
  Maroufy}}, \citenamefont {Kernane}, \citenamefont {Becheket},\ and\
  \citenamefont {Ouddadj}}]{ElMaroufy2016SIRBayesianInference}%
  \BibitemOpen
  \bibfield  {author} {\bibinfo {author} {\bibfnamefont {H.}~\bibnamefont {{El
  Maroufy}}}, \bibinfo {author} {\bibfnamefont {T.}~\bibnamefont {Kernane}},
  \bibinfo {author} {\bibfnamefont {S.}~\bibnamefont {Becheket}}, \ and\
  \bibinfo {author} {\bibfnamefont {A.}~\bibnamefont {Ouddadj}},\ }\href
  {\doibase 10.1080/00949655.2015.1107561} {\bibfield  {journal} {\bibinfo
  {journal} {Journal of Statistical Computation and Simulation}\ }\textbf
  {\bibinfo {volume} {86}},\ \bibinfo {pages} {2229} (\bibinfo {year}
  {2016})}\BibitemShut {NoStop}%
\bibitem [{\citenamefont {Shah}\ \emph {et~al.}(2020)\citenamefont {Shah},
  \citenamefont {Dehmamy}, \citenamefont {Perra}, \citenamefont {Chinazzi},
  \citenamefont {Barab{\'a}si}, \citenamefont {Vespignani},\ and\ \citenamefont
  {Yu}}]{shah2020finding}%
  \BibitemOpen
  \bibfield  {author} {\bibinfo {author} {\bibfnamefont {C.}~\bibnamefont
  {Shah}}, \bibinfo {author} {\bibfnamefont {N.}~\bibnamefont {Dehmamy}},
  \bibinfo {author} {\bibfnamefont {N.}~\bibnamefont {Perra}}, \bibinfo
  {author} {\bibfnamefont {M.}~\bibnamefont {Chinazzi}}, \bibinfo {author}
  {\bibfnamefont {A.-L.}\ \bibnamefont {Barab{\'a}si}}, \bibinfo {author}
  {\bibfnamefont {A.}~\bibnamefont {Vespignani}}, \ and\ \bibinfo {author}
  {\bibfnamefont {R.}~\bibnamefont {Yu}},\ }\href@noop {} {\bibfield  {journal}
  {\bibinfo  {journal} {arXiv preprint arXiv:2006.11913}\ } (\bibinfo {year}
  {2020})}\BibitemShut {NoStop}%
\bibitem [{\citenamefont {Shen}\ \emph {et~al.}(2016)\citenamefont {Shen},
  \citenamefont {Cao}, \citenamefont {Wang}, \citenamefont {Di},\ and\
  \citenamefont {Stanley}}]{shen2016locating}%
  \BibitemOpen
  \bibfield  {author} {\bibinfo {author} {\bibfnamefont {Z.}~\bibnamefont
  {Shen}}, \bibinfo {author} {\bibfnamefont {S.}~\bibnamefont {Cao}}, \bibinfo
  {author} {\bibfnamefont {W.-X.}\ \bibnamefont {Wang}}, \bibinfo {author}
  {\bibfnamefont {Z.}~\bibnamefont {Di}}, \ and\ \bibinfo {author}
  {\bibfnamefont {H.~E.}\ \bibnamefont {Stanley}},\ }\href@noop {} {\bibfield
  {journal} {\bibinfo  {journal} {Physical Review E}\ }\textbf {\bibinfo
  {volume} {93}},\ \bibinfo {pages} {032301} (\bibinfo {year}
  {2016})}\BibitemShut {NoStop}%
\bibitem [{\citenamefont {Pinto}\ \emph {et~al.}(2012)\citenamefont {Pinto},
  \citenamefont {Thiran},\ and\ \citenamefont {Vetterli}}]{pinto2012locating}%
  \BibitemOpen
  \bibfield  {author} {\bibinfo {author} {\bibfnamefont {P.~C.}\ \bibnamefont
  {Pinto}}, \bibinfo {author} {\bibfnamefont {P.}~\bibnamefont {Thiran}}, \
  and\ \bibinfo {author} {\bibfnamefont {M.}~\bibnamefont {Vetterli}},\
  }\href@noop {} {\bibfield  {journal} {\bibinfo  {journal} {Physical review
  letters}\ }\textbf {\bibinfo {volume} {109}},\ \bibinfo {pages} {068702}
  (\bibinfo {year} {2012})}\BibitemShut {NoStop}%
\bibitem [{\citenamefont {Marinelli}\ and\ \citenamefont
  {Faraldo-G{\'o}mez}(2015)}]{marinelli2015ensemble}%
  \BibitemOpen
  \bibfield  {author} {\bibinfo {author} {\bibfnamefont {F.}~\bibnamefont
  {Marinelli}}\ and\ \bibinfo {author} {\bibfnamefont {J.~D.}\ \bibnamefont
  {Faraldo-G{\'o}mez}},\ }\href@noop {} {\bibfield  {journal} {\bibinfo
  {journal} {Biophysical journal}\ }\textbf {\bibinfo {volume} {108}},\
  \bibinfo {pages} {2779} (\bibinfo {year} {2015})}\BibitemShut {NoStop}%
\bibitem [{\citenamefont {Cesari}\ \emph {et~al.}(2018)\citenamefont {Cesari},
  \citenamefont {Rei{\ss}er},\ and\ \citenamefont {Bussi}}]{cesari2018using}%
  \BibitemOpen
  \bibfield  {author} {\bibinfo {author} {\bibfnamefont {A.}~\bibnamefont
  {Cesari}}, \bibinfo {author} {\bibfnamefont {S.}~\bibnamefont {Rei{\ss}er}},
  \ and\ \bibinfo {author} {\bibfnamefont {G.}~\bibnamefont {Bussi}},\
  }\href@noop {} {\bibfield  {journal} {\bibinfo  {journal} {Computation}\
  }\textbf {\bibinfo {volume} {6}},\ \bibinfo {pages} {15} (\bibinfo {year}
  {2018})}\BibitemShut {NoStop}%
\bibitem [{\citenamefont {Amirkulova}\ and\ \citenamefont
  {White}(2019)}]{Amirkulova2019}%
  \BibitemOpen
  \bibfield  {author} {\bibinfo {author} {\bibfnamefont {D.~B.}\ \bibnamefont
  {Amirkulova}}\ and\ \bibinfo {author} {\bibfnamefont {A.~D.}\ \bibnamefont
  {White}},\ }\href {\doibase 10.1080/08927022.2019.1608988} {\bibfield
  {journal} {\bibinfo  {journal} {Molecular Simulation}\ }\textbf {\bibinfo
  {volume} {45}},\ \bibinfo {pages} {1285} (\bibinfo {year} {2019})},\ \Eprint
  {http://arxiv.org/abs/https://doi.org/10.1080/08927022.2019.1608988}
  {https://doi.org/10.1080/08927022.2019.1608988} \BibitemShut {NoStop}%
\bibitem [{\citenamefont {Shipley}\ \emph {et~al.}(2006)\citenamefont
  {Shipley}, \citenamefont {Vile},\ and\ \citenamefont
  {Garnier}}]{shipley2006plant}%
  \BibitemOpen
  \bibfield  {author} {\bibinfo {author} {\bibfnamefont {B.}~\bibnamefont
  {Shipley}}, \bibinfo {author} {\bibfnamefont {D.}~\bibnamefont {Vile}}, \
  and\ \bibinfo {author} {\bibfnamefont {{\'E}.}~\bibnamefont {Garnier}},\
  }\href@noop {} {\bibfield  {journal} {\bibinfo  {journal} {science}\ }\textbf
  {\bibinfo {volume} {314}},\ \bibinfo {pages} {812} (\bibinfo {year}
  {2006})}\BibitemShut {NoStop}%
\bibitem [{\citenamefont {Harte}\ \emph {et~al.}(2008)\citenamefont {Harte},
  \citenamefont {Zillio}, \citenamefont {Conlisk},\ and\ \citenamefont
  {Smith}}]{harte2008maximum}%
  \BibitemOpen
  \bibfield  {author} {\bibinfo {author} {\bibfnamefont {J.}~\bibnamefont
  {Harte}}, \bibinfo {author} {\bibfnamefont {T.}~\bibnamefont {Zillio}},
  \bibinfo {author} {\bibfnamefont {E.}~\bibnamefont {Conlisk}}, \ and\
  \bibinfo {author} {\bibfnamefont {A.~B.}\ \bibnamefont {Smith}},\ }\href@noop
  {} {\bibfield  {journal} {\bibinfo  {journal} {Ecology}\ }\textbf {\bibinfo
  {volume} {89}},\ \bibinfo {pages} {2700} (\bibinfo {year}
  {2008})}\BibitemShut {NoStop}%
\bibitem [{\citenamefont {Dewar}\ and\ \citenamefont
  {Port{\'e}}(2008)}]{dewar2008statistical}%
  \BibitemOpen
  \bibfield  {author} {\bibinfo {author} {\bibfnamefont {R.~C.}\ \bibnamefont
  {Dewar}}\ and\ \bibinfo {author} {\bibfnamefont {A.}~\bibnamefont
  {Port{\'e}}},\ }\href@noop {} {\bibfield  {journal} {\bibinfo  {journal}
  {Journal of theoretical biology}\ }\textbf {\bibinfo {volume} {251}},\
  \bibinfo {pages} {389} (\bibinfo {year} {2008})}\BibitemShut {NoStop}%
\bibitem [{\citenamefont {Favretti}(2018)}]{favretti2018remarks}%
  \BibitemOpen
  \bibfield  {author} {\bibinfo {author} {\bibfnamefont {M.}~\bibnamefont
  {Favretti}},\ }\href@noop {} {\bibfield  {journal} {\bibinfo  {journal}
  {Entropy}\ }\textbf {\bibinfo {volume} {20}},\ \bibinfo {pages} {11}
  (\bibinfo {year} {2018})}\BibitemShut {NoStop}%
\bibitem [{\citenamefont {Sibisi}\ \emph {et~al.}(1984)\citenamefont {Sibisi},
  \citenamefont {Skilling}, \citenamefont {Brereton}, \citenamefont {Laue},\
  and\ \citenamefont {Staunton}}]{sibisi1984maximum}%
  \BibitemOpen
  \bibfield  {author} {\bibinfo {author} {\bibfnamefont {S.}~\bibnamefont
  {Sibisi}}, \bibinfo {author} {\bibfnamefont {J.}~\bibnamefont {Skilling}},
  \bibinfo {author} {\bibfnamefont {R.~G.}\ \bibnamefont {Brereton}}, \bibinfo
  {author} {\bibfnamefont {E.~D.}\ \bibnamefont {Laue}}, \ and\ \bibinfo
  {author} {\bibfnamefont {J.}~\bibnamefont {Staunton}},\ }\href@noop {}
  {\bibfield  {journal} {\bibinfo  {journal} {Nature}\ }\textbf {\bibinfo
  {volume} {311}},\ \bibinfo {pages} {446} (\bibinfo {year}
  {1984})}\BibitemShut {NoStop}%
\bibitem [{\citenamefont {Hoch}\ \emph {et~al.}(2014)\citenamefont {Hoch},
  \citenamefont {Maciejewski}, \citenamefont {Mobli}, \citenamefont
  {Schuyler},\ and\ \citenamefont {Stern}}]{hoch2014nonuniform}%
  \BibitemOpen
  \bibfield  {author} {\bibinfo {author} {\bibfnamefont {J.~C.}\ \bibnamefont
  {Hoch}}, \bibinfo {author} {\bibfnamefont {M.~W.}\ \bibnamefont
  {Maciejewski}}, \bibinfo {author} {\bibfnamefont {M.}~\bibnamefont {Mobli}},
  \bibinfo {author} {\bibfnamefont {A.~D.}\ \bibnamefont {Schuyler}}, \ and\
  \bibinfo {author} {\bibfnamefont {A.~S.}\ \bibnamefont {Stern}},\ }\href@noop
  {} {\bibfield  {journal} {\bibinfo  {journal} {Accounts of chemical
  research}\ }\textbf {\bibinfo {volume} {47}},\ \bibinfo {pages} {708}
  (\bibinfo {year} {2014})}\BibitemShut {NoStop}%
\bibitem [{\citenamefont {Kitaura}\ \emph {et~al.}(2002)\citenamefont
  {Kitaura}, \citenamefont {Kitagawa}, \citenamefont {Kubota}, \citenamefont
  {Kobayashi}, \citenamefont {Kindo}, \citenamefont {Mita}, \citenamefont
  {Matsuo}, \citenamefont {Kobayashi}, \citenamefont {Chang}, \citenamefont
  {Ozawa} \emph {et~al.}}]{kitaura2002formation}%
  \BibitemOpen
  \bibfield  {author} {\bibinfo {author} {\bibfnamefont {R.}~\bibnamefont
  {Kitaura}}, \bibinfo {author} {\bibfnamefont {S.}~\bibnamefont {Kitagawa}},
  \bibinfo {author} {\bibfnamefont {Y.}~\bibnamefont {Kubota}}, \bibinfo
  {author} {\bibfnamefont {T.~C.}\ \bibnamefont {Kobayashi}}, \bibinfo {author}
  {\bibfnamefont {K.}~\bibnamefont {Kindo}}, \bibinfo {author} {\bibfnamefont
  {Y.}~\bibnamefont {Mita}}, \bibinfo {author} {\bibfnamefont {A.}~\bibnamefont
  {Matsuo}}, \bibinfo {author} {\bibfnamefont {M.}~\bibnamefont {Kobayashi}},
  \bibinfo {author} {\bibfnamefont {H.-C.}\ \bibnamefont {Chang}}, \bibinfo
  {author} {\bibfnamefont {T.~C.}\ \bibnamefont {Ozawa}},  \emph {et~al.},\
  }\href@noop {} {\bibfield  {journal} {\bibinfo  {journal} {Science}\ }\textbf
  {\bibinfo {volume} {298}},\ \bibinfo {pages} {2358} (\bibinfo {year}
  {2002})}\BibitemShut {NoStop}%
\bibitem [{\citenamefont {Andersen}\ \emph {et~al.}(2014)\citenamefont
  {Andersen}, \citenamefont {Bremholm}, \citenamefont {Vennestr{\o}m},
  \citenamefont {Blichfeld}, \citenamefont {Lundegaard},\ and\ \citenamefont
  {Iversen}}]{andersen2014location}%
  \BibitemOpen
  \bibfield  {author} {\bibinfo {author} {\bibfnamefont {C.~W.}\ \bibnamefont
  {Andersen}}, \bibinfo {author} {\bibfnamefont {M.}~\bibnamefont {Bremholm}},
  \bibinfo {author} {\bibfnamefont {P.~N.~R.}\ \bibnamefont {Vennestr{\o}m}},
  \bibinfo {author} {\bibfnamefont {A.~B.}\ \bibnamefont {Blichfeld}}, \bibinfo
  {author} {\bibfnamefont {L.~F.}\ \bibnamefont {Lundegaard}}, \ and\ \bibinfo
  {author} {\bibfnamefont {B.~B.}\ \bibnamefont {Iversen}},\ }\href@noop {}
  {\bibfield  {journal} {\bibinfo  {journal} {IUCrJ}\ }\textbf {\bibinfo
  {volume} {1}},\ \bibinfo {pages} {382} (\bibinfo {year} {2014})}\BibitemShut
  {NoStop}%
\bibitem [{\citenamefont {Ferrige}\ \emph {et~al.}(1992)\citenamefont
  {Ferrige}, \citenamefont {Seddon}, \citenamefont {Jarvis}, \citenamefont
  {Skilling},\ and\ \citenamefont {Welch}}]{ferrige1992application}%
  \BibitemOpen
  \bibfield  {author} {\bibinfo {author} {\bibfnamefont {A.}~\bibnamefont
  {Ferrige}}, \bibinfo {author} {\bibfnamefont {M.}~\bibnamefont {Seddon}},
  \bibinfo {author} {\bibfnamefont {S.}~\bibnamefont {Jarvis}}, \bibinfo
  {author} {\bibfnamefont {J.}~\bibnamefont {Skilling}}, \ and\ \bibinfo
  {author} {\bibfnamefont {J.}~\bibnamefont {Welch}},\ }in\ \href@noop {}
  {\emph {\bibinfo {booktitle} {Maximum Entropy and Bayesian Methods}}}\
  (\bibinfo  {publisher} {Springer},\ \bibinfo {year} {1992})\ pp.\ \bibinfo
  {pages} {327--335}\BibitemShut {NoStop}%
\bibitem [{\citenamefont {Kimoto}\ \emph {et~al.}(2010)\citenamefont {Kimoto},
  \citenamefont {Asaka}, \citenamefont {Yu}, \citenamefont {Nagai},
  \citenamefont {Matsui},\ and\ \citenamefont {Ishizuka}}]{kimoto2010local}%
  \BibitemOpen
  \bibfield  {author} {\bibinfo {author} {\bibfnamefont {K.}~\bibnamefont
  {Kimoto}}, \bibinfo {author} {\bibfnamefont {T.}~\bibnamefont {Asaka}},
  \bibinfo {author} {\bibfnamefont {X.}~\bibnamefont {Yu}}, \bibinfo {author}
  {\bibfnamefont {T.}~\bibnamefont {Nagai}}, \bibinfo {author} {\bibfnamefont
  {Y.}~\bibnamefont {Matsui}}, \ and\ \bibinfo {author} {\bibfnamefont
  {K.}~\bibnamefont {Ishizuka}},\ }\href@noop {} {\bibfield  {journal}
  {\bibinfo  {journal} {Ultramicroscopy}\ }\textbf {\bibinfo {volume} {110}},\
  \bibinfo {pages} {778} (\bibinfo {year} {2010})}\BibitemShut {NoStop}%
\bibitem [{\citenamefont {Scharfenaker}\ and\ \citenamefont
  {Yang}(2020)}]{scharfenaker2020maximum}%
  \BibitemOpen
  \bibfield  {author} {\bibinfo {author} {\bibfnamefont {E.}~\bibnamefont
  {Scharfenaker}}\ and\ \bibinfo {author} {\bibfnamefont {J.}~\bibnamefont
  {Yang}},\ }\href@noop {} {\bibfield  {journal} {\bibinfo  {journal} {The
  European Physical Journal Special Topics}\ }\textbf {\bibinfo {volume}
  {229}},\ \bibinfo {pages} {1577} (\bibinfo {year} {2020})}\BibitemShut
  {NoStop}%
\bibitem [{\citenamefont {Schneidman}\ \emph {et~al.}(2006)\citenamefont
  {Schneidman}, \citenamefont {Berry}, \citenamefont {Segev},\ and\
  \citenamefont {Bialek}}]{schneidman2006weak}%
  \BibitemOpen
  \bibfield  {author} {\bibinfo {author} {\bibfnamefont {E.}~\bibnamefont
  {Schneidman}}, \bibinfo {author} {\bibfnamefont {M.~J.}\ \bibnamefont
  {Berry}}, \bibinfo {author} {\bibfnamefont {R.}~\bibnamefont {Segev}}, \ and\
  \bibinfo {author} {\bibfnamefont {W.}~\bibnamefont {Bialek}},\ }\href@noop {}
  {\bibfield  {journal} {\bibinfo  {journal} {Nature}\ }\textbf {\bibinfo
  {volume} {440}},\ \bibinfo {pages} {1007} (\bibinfo {year}
  {2006})}\BibitemShut {NoStop}%
\bibitem [{\citenamefont {Tang}\ \emph {et~al.}(2008)\citenamefont {Tang},
  \citenamefont {Jackson}, \citenamefont {Hobbs}, \citenamefont {Chen},
  \citenamefont {Smith}, \citenamefont {Patel}, \citenamefont {Prieto},
  \citenamefont {Petrusca}, \citenamefont {Grivich}, \citenamefont {Sher} \emph
  {et~al.}}]{tang2008maximum}%
  \BibitemOpen
  \bibfield  {author} {\bibinfo {author} {\bibfnamefont {A.}~\bibnamefont
  {Tang}}, \bibinfo {author} {\bibfnamefont {D.}~\bibnamefont {Jackson}},
  \bibinfo {author} {\bibfnamefont {J.}~\bibnamefont {Hobbs}}, \bibinfo
  {author} {\bibfnamefont {W.}~\bibnamefont {Chen}}, \bibinfo {author}
  {\bibfnamefont {J.~L.}\ \bibnamefont {Smith}}, \bibinfo {author}
  {\bibfnamefont {H.}~\bibnamefont {Patel}}, \bibinfo {author} {\bibfnamefont
  {A.}~\bibnamefont {Prieto}}, \bibinfo {author} {\bibfnamefont
  {D.}~\bibnamefont {Petrusca}}, \bibinfo {author} {\bibfnamefont {M.~I.}\
  \bibnamefont {Grivich}}, \bibinfo {author} {\bibfnamefont {A.}~\bibnamefont
  {Sher}},  \emph {et~al.},\ }\href@noop {} {\bibfield  {journal} {\bibinfo
  {journal} {Journal of Neuroscience}\ }\textbf {\bibinfo {volume} {28}},\
  \bibinfo {pages} {505} (\bibinfo {year} {2008})}\BibitemShut {NoStop}%
\bibitem [{\citenamefont {Granot-Atedgi}\ \emph {et~al.}(2013)\citenamefont
  {Granot-Atedgi}, \citenamefont {Tka{\v{c}}ik}, \citenamefont {Segev},\ and\
  \citenamefont {Schneidman}}]{granot2013stimulus}%
  \BibitemOpen
  \bibfield  {author} {\bibinfo {author} {\bibfnamefont {E.}~\bibnamefont
  {Granot-Atedgi}}, \bibinfo {author} {\bibfnamefont {G.}~\bibnamefont
  {Tka{\v{c}}ik}}, \bibinfo {author} {\bibfnamefont {R.}~\bibnamefont {Segev}},
  \ and\ \bibinfo {author} {\bibfnamefont {E.}~\bibnamefont {Schneidman}},\
  }\href@noop {} {\bibfield  {journal} {\bibinfo  {journal} {PLoS computational
  biology}\ }\textbf {\bibinfo {volume} {9}},\ \bibinfo {pages} {e1002922}
  (\bibinfo {year} {2013})}\BibitemShut {NoStop}%
\bibitem [{\citenamefont {Watanabe}\ \emph {et~al.}(2013)\citenamefont
  {Watanabe}, \citenamefont {Hirose}, \citenamefont {Wada}, \citenamefont
  {Imai}, \citenamefont {Machida}, \citenamefont {Shirouzu}, \citenamefont
  {Konishi}, \citenamefont {Miyashita},\ and\ \citenamefont
  {Masuda}}]{watanabe2013pairwise}%
  \BibitemOpen
  \bibfield  {author} {\bibinfo {author} {\bibfnamefont {T.}~\bibnamefont
  {Watanabe}}, \bibinfo {author} {\bibfnamefont {S.}~\bibnamefont {Hirose}},
  \bibinfo {author} {\bibfnamefont {H.}~\bibnamefont {Wada}}, \bibinfo {author}
  {\bibfnamefont {Y.}~\bibnamefont {Imai}}, \bibinfo {author} {\bibfnamefont
  {T.}~\bibnamefont {Machida}}, \bibinfo {author} {\bibfnamefont
  {I.}~\bibnamefont {Shirouzu}}, \bibinfo {author} {\bibfnamefont
  {S.}~\bibnamefont {Konishi}}, \bibinfo {author} {\bibfnamefont
  {Y.}~\bibnamefont {Miyashita}}, \ and\ \bibinfo {author} {\bibfnamefont
  {N.}~\bibnamefont {Masuda}},\ }\href@noop {} {\bibfield  {journal} {\bibinfo
  {journal} {Nature communications}\ }\textbf {\bibinfo {volume} {4}},\
  \bibinfo {pages} {1} (\bibinfo {year} {2013})}\BibitemShut {NoStop}%
\bibitem [{\citenamefont {Barrett}\ \emph {et~al.}(2021)\citenamefont
  {Barrett}, \citenamefont {Ansari}, \citenamefont {Ghoshal},\ and\
  \citenamefont {White}}]{barrett2021simulation}%
  \BibitemOpen
  \bibfield  {author} {\bibinfo {author} {\bibfnamefont {R.}~\bibnamefont
  {Barrett}}, \bibinfo {author} {\bibfnamefont {M.}~\bibnamefont {Ansari}},
  \bibinfo {author} {\bibfnamefont {G.}~\bibnamefont {Ghoshal}}, \ and\
  \bibinfo {author} {\bibfnamefont {A.~D.}\ \bibnamefont {White}},\ }\href@noop
  {} {\bibfield  {journal} {\bibinfo  {journal} {arXiv preprint
  arXiv:2104.09668}\ } (\bibinfo {year} {2021})}\BibitemShut {NoStop}%
\bibitem [{\citenamefont {Vespignani}\ \emph {et~al.}(2020)\citenamefont
  {Vespignani}, \citenamefont {Tian}, \citenamefont {Dye}, \citenamefont
  {Lloyd-Smith}, \citenamefont {Eggo}, \citenamefont {Shrestha}, \citenamefont
  {Scarpino}, \citenamefont {Gutierrez}, \citenamefont {Kraemer}, \citenamefont
  {Wu} \emph {et~al.}}]{vespignani2020modelling}%
  \BibitemOpen
  \bibfield  {author} {\bibinfo {author} {\bibfnamefont {A.}~\bibnamefont
  {Vespignani}}, \bibinfo {author} {\bibfnamefont {H.}~\bibnamefont {Tian}},
  \bibinfo {author} {\bibfnamefont {C.}~\bibnamefont {Dye}}, \bibinfo {author}
  {\bibfnamefont {J.~O.}\ \bibnamefont {Lloyd-Smith}}, \bibinfo {author}
  {\bibfnamefont {R.~M.}\ \bibnamefont {Eggo}}, \bibinfo {author}
  {\bibfnamefont {M.}~\bibnamefont {Shrestha}}, \bibinfo {author}
  {\bibfnamefont {S.~V.}\ \bibnamefont {Scarpino}}, \bibinfo {author}
  {\bibfnamefont {B.}~\bibnamefont {Gutierrez}}, \bibinfo {author}
  {\bibfnamefont {M.~U.}\ \bibnamefont {Kraemer}}, \bibinfo {author}
  {\bibfnamefont {J.}~\bibnamefont {Wu}},  \emph {et~al.},\ }\href@noop {}
  {\bibfield  {journal} {\bibinfo  {journal} {Nature Reviews Physics}\ }\textbf
  {\bibinfo {volume} {2}},\ \bibinfo {pages} {279} (\bibinfo {year}
  {2020})}\BibitemShut {NoStop}%
\bibitem [{\citenamefont {Artalejo}\ and\ \citenamefont
  {Lopez-Herrero}(2011)}]{artalejo2011sis}%
  \BibitemOpen
  \bibfield  {author} {\bibinfo {author} {\bibfnamefont {J.~R.}\ \bibnamefont
  {Artalejo}}\ and\ \bibinfo {author} {\bibfnamefont {M.}~\bibnamefont
  {Lopez-Herrero}},\ }\href@noop {} {\bibfield  {journal} {\bibinfo  {journal}
  {Theoretical population biology}\ }\textbf {\bibinfo {volume} {80}},\
  \bibinfo {pages} {256} (\bibinfo {year} {2011})}\BibitemShut {NoStop}%
\bibitem [{\citenamefont {Harding}\ \emph {et~al.}(2020)\citenamefont
  {Harding}, \citenamefont {Spinney},\ and\ \citenamefont
  {Prokopenko}}]{harding2020population}%
  \BibitemOpen
  \bibfield  {author} {\bibinfo {author} {\bibfnamefont {N.}~\bibnamefont
  {Harding}}, \bibinfo {author} {\bibfnamefont {R.~E.}\ \bibnamefont
  {Spinney}}, \ and\ \bibinfo {author} {\bibfnamefont {M.}~\bibnamefont
  {Prokopenko}},\ }\href@noop {} {\bibfield  {journal} {\bibinfo  {journal}
  {Scientific reports}\ }\textbf {\bibinfo {volume} {10}},\ \bibinfo {pages}
  {1} (\bibinfo {year} {2020})}\BibitemShut {NoStop}%
\bibitem [{\citenamefont {Hanski}(1998)}]{hanski1998metapopulation}%
  \BibitemOpen
  \bibfield  {author} {\bibinfo {author} {\bibfnamefont {I.}~\bibnamefont
  {Hanski}},\ }\href@noop {} {\bibfield  {journal} {\bibinfo  {journal}
  {Nature}\ }\textbf {\bibinfo {volume} {396}},\ \bibinfo {pages} {41}
  (\bibinfo {year} {1998})}\BibitemShut {NoStop}%
\bibitem [{\citenamefont {Ball}\ \emph {et~al.}(2015)\citenamefont {Ball},
  \citenamefont {Britton}, \citenamefont {House}, \citenamefont {Isham},
  \citenamefont {Mollison}, \citenamefont {Pellis},\ and\ \citenamefont
  {Tomba}}]{ball2015seven}%
  \BibitemOpen
  \bibfield  {author} {\bibinfo {author} {\bibfnamefont {F.}~\bibnamefont
  {Ball}}, \bibinfo {author} {\bibfnamefont {T.}~\bibnamefont {Britton}},
  \bibinfo {author} {\bibfnamefont {T.}~\bibnamefont {House}}, \bibinfo
  {author} {\bibfnamefont {V.}~\bibnamefont {Isham}}, \bibinfo {author}
  {\bibfnamefont {D.}~\bibnamefont {Mollison}}, \bibinfo {author}
  {\bibfnamefont {L.}~\bibnamefont {Pellis}}, \ and\ \bibinfo {author}
  {\bibfnamefont {G.~S.}\ \bibnamefont {Tomba}},\ }\href@noop {} {\bibfield
  {journal} {\bibinfo  {journal} {Epidemics}\ }\textbf {\bibinfo {volume}
  {10}},\ \bibinfo {pages} {63} (\bibinfo {year} {2015})}\BibitemShut {NoStop}%
\bibitem [{\citenamefont {Hagberg}\ \emph {et~al.}(2008)\citenamefont
  {Hagberg}, \citenamefont {Swart},\ and\ \citenamefont
  {S~Chult}}]{hagberg2008exploring}%
  \BibitemOpen
  \bibfield  {author} {\bibinfo {author} {\bibfnamefont {A.}~\bibnamefont
  {Hagberg}}, \bibinfo {author} {\bibfnamefont {P.}~\bibnamefont {Swart}}, \
  and\ \bibinfo {author} {\bibfnamefont {D.}~\bibnamefont {S~Chult}},\
  }\href@noop {} {\emph {\bibinfo {title} {Exploring network structure,
  dynamics, and function using NetworkX}}},\ \bibinfo {type} {Tech. Rep.}\
  (\bibinfo  {institution} {Los Alamos National Lab.(LANL), Los Alamos, NM
  (United States)},\ \bibinfo {year} {2008})\BibitemShut {NoStop}%
\bibitem [{\citenamefont {Roux}\ and\ \citenamefont
  {Weare}(2013)}]{Roux2013statistical}%
  \BibitemOpen
  \bibfield  {author} {\bibinfo {author} {\bibfnamefont {B.}~\bibnamefont
  {Roux}}\ and\ \bibinfo {author} {\bibfnamefont {J.}~\bibnamefont {Weare}},\
  }\href@noop {} {\bibfield  {journal} {\bibinfo  {journal} {The Journal of
  chemical physics}\ }\textbf {\bibinfo {volume} {138}},\ \bibinfo {pages}
  {02B616} (\bibinfo {year} {2013})}\BibitemShut {NoStop}%
\bibitem [{\citenamefont {Pitera}\ and\ \citenamefont
  {Chodera}(2012)}]{Pitera2012}%
  \BibitemOpen
  \bibfield  {author} {\bibinfo {author} {\bibfnamefont {J.~W.}\ \bibnamefont
  {Pitera}}\ and\ \bibinfo {author} {\bibfnamefont {J.~D.}\ \bibnamefont
  {Chodera}},\ }\href {\doibase 10.1021/ct300112v} {\bibfield  {journal}
  {\bibinfo  {journal} {Journal of Chemical Theory and Computation}\ }\textbf
  {\bibinfo {volume} {8}},\ \bibinfo {pages} {3445} (\bibinfo {year}
  {2012})}\BibitemShut {NoStop}%
\bibitem [{\citenamefont {Cesari}\ \emph {et~al.}(2016)\citenamefont {Cesari},
  \citenamefont {Gil-Ley},\ and\ \citenamefont
  {Bussi}}]{cesari2016MaxEntUncertainty}%
  \BibitemOpen
  \bibfield  {author} {\bibinfo {author} {\bibfnamefont {A.}~\bibnamefont
  {Cesari}}, \bibinfo {author} {\bibfnamefont {A.}~\bibnamefont {Gil-Ley}}, \
  and\ \bibinfo {author} {\bibfnamefont {G.}~\bibnamefont {Bussi}},\
  }\href@noop {} {\bibfield  {journal} {\bibinfo  {journal} {Journal of
  chemical theory and computation}\ }\textbf {\bibinfo {volume} {12}},\
  \bibinfo {pages} {6192} (\bibinfo {year} {2016})}\BibitemShut {NoStop}%
\bibitem [{\citenamefont {Grenfell}\ and\ \citenamefont
  {Harwood}(1997)}]{grenfell1997meta}%
  \BibitemOpen
  \bibfield  {author} {\bibinfo {author} {\bibfnamefont {B.}~\bibnamefont
  {Grenfell}}\ and\ \bibinfo {author} {\bibfnamefont {J.}~\bibnamefont
  {Harwood}},\ }\href@noop {} {\bibfield  {journal} {\bibinfo  {journal}
  {Trends in ecology \& evolution}\ }\textbf {\bibinfo {volume} {12}},\
  \bibinfo {pages} {395} (\bibinfo {year} {1997})}\BibitemShut {NoStop}%
\bibitem [{\citenamefont {Watts}\ \emph {et~al.}(2005)\citenamefont {Watts},
  \citenamefont {Muhamad}, \citenamefont {Medina},\ and\ \citenamefont
  {Dodds}}]{watts2005multiscale}%
  \BibitemOpen
  \bibfield  {author} {\bibinfo {author} {\bibfnamefont {D.~J.}\ \bibnamefont
  {Watts}}, \bibinfo {author} {\bibfnamefont {R.}~\bibnamefont {Muhamad}},
  \bibinfo {author} {\bibfnamefont {D.~C.}\ \bibnamefont {Medina}}, \ and\
  \bibinfo {author} {\bibfnamefont {P.~S.}\ \bibnamefont {Dodds}},\ }\href@noop
  {} {\bibfield  {journal} {\bibinfo  {journal} {Proceedings of the National
  Academy of Sciences}\ }\textbf {\bibinfo {volume} {102}},\ \bibinfo {pages}
  {11157} (\bibinfo {year} {2005})}\BibitemShut {NoStop}%
\bibitem [{\citenamefont {Colizza}\ \emph {et~al.}(2007)\citenamefont
  {Colizza}, \citenamefont {Pastor-Satorras},\ and\ \citenamefont
  {Vespignani}}]{colizza2007reaction}%
  \BibitemOpen
  \bibfield  {author} {\bibinfo {author} {\bibfnamefont {V.}~\bibnamefont
  {Colizza}}, \bibinfo {author} {\bibfnamefont {R.}~\bibnamefont
  {Pastor-Satorras}}, \ and\ \bibinfo {author} {\bibfnamefont {A.}~\bibnamefont
  {Vespignani}},\ }\href@noop {} {\bibfield  {journal} {\bibinfo  {journal}
  {Nature Physics}\ }\textbf {\bibinfo {volume} {3}},\ \bibinfo {pages} {276}
  (\bibinfo {year} {2007})}\BibitemShut {NoStop}%
\bibitem [{\citenamefont {Zhou}\ \emph {et~al.}(2020)\citenamefont {Zhou},
  \citenamefont {Yang}, \citenamefont {Wang}, \citenamefont {Hu}, \citenamefont
  {Zhang}, \citenamefont {Zhang}, \citenamefont {Si}, \citenamefont {Zhu},
  \citenamefont {Li}, \citenamefont {Huang} \emph
  {et~al.}}]{zhou2020pneumonia}%
  \BibitemOpen
  \bibfield  {author} {\bibinfo {author} {\bibfnamefont {P.}~\bibnamefont
  {Zhou}}, \bibinfo {author} {\bibfnamefont {X.-L.}\ \bibnamefont {Yang}},
  \bibinfo {author} {\bibfnamefont {X.-G.}\ \bibnamefont {Wang}}, \bibinfo
  {author} {\bibfnamefont {B.}~\bibnamefont {Hu}}, \bibinfo {author}
  {\bibfnamefont {L.}~\bibnamefont {Zhang}}, \bibinfo {author} {\bibfnamefont
  {W.}~\bibnamefont {Zhang}}, \bibinfo {author} {\bibfnamefont {H.-R.}\
  \bibnamefont {Si}}, \bibinfo {author} {\bibfnamefont {Y.}~\bibnamefont
  {Zhu}}, \bibinfo {author} {\bibfnamefont {B.}~\bibnamefont {Li}}, \bibinfo
  {author} {\bibfnamefont {C.-L.}\ \bibnamefont {Huang}},  \emph {et~al.},\
  }\href@noop {} {\bibfield  {journal} {\bibinfo  {journal} {nature}\ }\textbf
  {\bibinfo {volume} {579}},\ \bibinfo {pages} {270} (\bibinfo {year}
  {2020})}\BibitemShut {NoStop}%
\bibitem [{\citenamefont {Wu}\ \emph {et~al.}(2020)\citenamefont {Wu},
  \citenamefont {Zhao}, \citenamefont {Yu}, \citenamefont {Chen}, \citenamefont
  {Wang}, \citenamefont {Song}, \citenamefont {Hu}, \citenamefont {Tao},
  \citenamefont {Tian}, \citenamefont {Pei} \emph {et~al.}}]{wu2020new}%
  \BibitemOpen
  \bibfield  {author} {\bibinfo {author} {\bibfnamefont {F.}~\bibnamefont
  {Wu}}, \bibinfo {author} {\bibfnamefont {S.}~\bibnamefont {Zhao}}, \bibinfo
  {author} {\bibfnamefont {B.}~\bibnamefont {Yu}}, \bibinfo {author}
  {\bibfnamefont {Y.-M.}\ \bibnamefont {Chen}}, \bibinfo {author}
  {\bibfnamefont {W.}~\bibnamefont {Wang}}, \bibinfo {author} {\bibfnamefont
  {Z.-G.}\ \bibnamefont {Song}}, \bibinfo {author} {\bibfnamefont
  {Y.}~\bibnamefont {Hu}}, \bibinfo {author} {\bibfnamefont {Z.-W.}\
  \bibnamefont {Tao}}, \bibinfo {author} {\bibfnamefont {J.-H.}\ \bibnamefont
  {Tian}}, \bibinfo {author} {\bibfnamefont {Y.-Y.}\ \bibnamefont {Pei}},
  \emph {et~al.},\ }\href@noop {} {\bibfield  {journal} {\bibinfo  {journal}
  {Nature}\ }\textbf {\bibinfo {volume} {579}},\ \bibinfo {pages} {265}
  (\bibinfo {year} {2020})}\BibitemShut {NoStop}%
\bibitem [{\citenamefont {G{\'o}mez-Gardenes}\ \emph
  {et~al.}(2018)\citenamefont {G{\'o}mez-Gardenes}, \citenamefont
  {Soriano-Panos},\ and\ \citenamefont {Arenas}}]{gomez2018critical}%
  \BibitemOpen
  \bibfield  {author} {\bibinfo {author} {\bibfnamefont {J.}~\bibnamefont
  {G{\'o}mez-Gardenes}}, \bibinfo {author} {\bibfnamefont {D.}~\bibnamefont
  {Soriano-Panos}}, \ and\ \bibinfo {author} {\bibfnamefont {A.}~\bibnamefont
  {Arenas}},\ }\href@noop {} {\bibfield  {journal} {\bibinfo  {journal} {Nature
  Physics}\ }\textbf {\bibinfo {volume} {14}},\ \bibinfo {pages} {391}
  (\bibinfo {year} {2018})}\BibitemShut {NoStop}%
\bibitem [{\citenamefont {Arenas}\ \emph {et~al.}(2020)\citenamefont {Arenas},
  \citenamefont {Cota}, \citenamefont {G{\'o}mez-Garde{\~n}es}, \citenamefont
  {G{\'o}mez}, \citenamefont {Granell}, \citenamefont {Matamalas},
  \citenamefont {Soriano-Pa{\~n}os},\ and\ \citenamefont
  {Steinegger}}]{arenas2020mathematical}%
  \BibitemOpen
  \bibfield  {author} {\bibinfo {author} {\bibfnamefont {A.}~\bibnamefont
  {Arenas}}, \bibinfo {author} {\bibfnamefont {W.}~\bibnamefont {Cota}},
  \bibinfo {author} {\bibfnamefont {J.}~\bibnamefont {G{\'o}mez-Garde{\~n}es}},
  \bibinfo {author} {\bibfnamefont {S.}~\bibnamefont {G{\'o}mez}}, \bibinfo
  {author} {\bibfnamefont {C.}~\bibnamefont {Granell}}, \bibinfo {author}
  {\bibfnamefont {J.~T.}\ \bibnamefont {Matamalas}}, \bibinfo {author}
  {\bibfnamefont {D.}~\bibnamefont {Soriano-Pa{\~n}os}}, \ and\ \bibinfo
  {author} {\bibfnamefont {B.}~\bibnamefont {Steinegger}},\ }\href@noop {}
  {\bibfield  {journal} {\bibinfo  {journal} {Physical Review X}\ }\textbf
  {\bibinfo {volume} {10}},\ \bibinfo {pages} {041055} (\bibinfo {year}
  {2020})}\BibitemShut {NoStop}%
\bibitem [{\citenamefont {Lau}\ \emph {et~al.}(2021)\citenamefont {Lau},
  \citenamefont {Khosrawipour}, \citenamefont {Kocbach}, \citenamefont {Ichii},
  \citenamefont {Bania},\ and\ \citenamefont
  {Khosrawipour}}]{lau2021evaluating}%
  \BibitemOpen
  \bibfield  {author} {\bibinfo {author} {\bibfnamefont {H.}~\bibnamefont
  {Lau}}, \bibinfo {author} {\bibfnamefont {T.}~\bibnamefont {Khosrawipour}},
  \bibinfo {author} {\bibfnamefont {P.}~\bibnamefont {Kocbach}}, \bibinfo
  {author} {\bibfnamefont {H.}~\bibnamefont {Ichii}}, \bibinfo {author}
  {\bibfnamefont {J.}~\bibnamefont {Bania}}, \ and\ \bibinfo {author}
  {\bibfnamefont {V.}~\bibnamefont {Khosrawipour}},\ }\href@noop {} {\bibfield
  {journal} {\bibinfo  {journal} {Pulmonology}\ }\textbf {\bibinfo {volume}
  {27}},\ \bibinfo {pages} {110} (\bibinfo {year} {2021})}\BibitemShut
  {NoStop}%
\bibitem [{\citenamefont {Kingma}\ and\ \citenamefont
  {Ba}(2014)}]{kingma2014adam}%
  \BibitemOpen
  \bibfield  {author} {\bibinfo {author} {\bibfnamefont {D.~P.}\ \bibnamefont
  {Kingma}}\ and\ \bibinfo {author} {\bibfnamefont {J.}~\bibnamefont {Ba}},\
  }\href@noop {} {\bibfield  {journal} {\bibinfo  {journal} {arXiv preprint
  arXiv:1412.6980}\ } (\bibinfo {year} {2014})}\BibitemShut {NoStop}%
\bibitem [{cen()}]{census_data}%
  \BibitemOpen
  \href@noop {} {\enquote {\bibinfo {title} {Census mobility data},}\ }\bibinfo
  {howpublished}
  {\url{https://lehd.ces.census.gov/data/lodes/LODES7/}}\BibitemShut {NoStop}%
\bibitem [{\citenamefont {Stroock}\ and\ \citenamefont
  {Varadhan}(2007)}]{stroock2007multidimensional}%
  \BibitemOpen
  \bibfield  {author} {\bibinfo {author} {\bibfnamefont {D.~W.}\ \bibnamefont
  {Stroock}}\ and\ \bibinfo {author} {\bibfnamefont {S.~S.}\ \bibnamefont
  {Varadhan}},\ }\href@noop {} {\emph {\bibinfo {title} {Multidimensional
  diffusion processes}}}\ (\bibinfo  {publisher} {Springer},\ \bibinfo {year}
  {2007})\BibitemShut {NoStop}%
\bibitem [{\citenamefont {Pastor-Satorras}\ \emph {et~al.}(2015)\citenamefont
  {Pastor-Satorras}, \citenamefont {Castellano}, \citenamefont {Van~Mieghem},\
  and\ \citenamefont {Vespignani}}]{pastor2015epidemic}%
  \BibitemOpen
  \bibfield  {author} {\bibinfo {author} {\bibfnamefont {R.}~\bibnamefont
  {Pastor-Satorras}}, \bibinfo {author} {\bibfnamefont {C.}~\bibnamefont
  {Castellano}}, \bibinfo {author} {\bibfnamefont {P.}~\bibnamefont
  {Van~Mieghem}}, \ and\ \bibinfo {author} {\bibfnamefont {A.}~\bibnamefont
  {Vespignani}},\ }\href@noop {} {\bibfield  {journal} {\bibinfo  {journal}
  {Reviews of modern physics}\ }\textbf {\bibinfo {volume} {87}},\ \bibinfo
  {pages} {925} (\bibinfo {year} {2015})}\BibitemShut {NoStop}%
\bibitem [{\citenamefont {Centola}\ and\ \citenamefont
  {Macy}(2007)}]{centola2007complex}%
  \BibitemOpen
  \bibfield  {author} {\bibinfo {author} {\bibfnamefont {D.}~\bibnamefont
  {Centola}}\ and\ \bibinfo {author} {\bibfnamefont {M.}~\bibnamefont {Macy}},\
  }\href@noop {} {\bibfield  {journal} {\bibinfo  {journal} {American journal
  of Sociology}\ }\textbf {\bibinfo {volume} {113}},\ \bibinfo {pages} {702}
  (\bibinfo {year} {2007})}\BibitemShut {NoStop}%
\bibitem [{\citenamefont {Baronchelli}(2018)}]{baronchelli2018emergence}%
  \BibitemOpen
  \bibfield  {author} {\bibinfo {author} {\bibfnamefont {A.}~\bibnamefont
  {Baronchelli}},\ }\href@noop {} {\bibfield  {journal} {\bibinfo  {journal}
  {Royal Society open science}\ }\textbf {\bibinfo {volume} {5}},\ \bibinfo
  {pages} {172189} (\bibinfo {year} {2018})}\BibitemShut {NoStop}%
\bibitem [{\citenamefont {Wang}\ \emph {et~al.}(2013)\citenamefont {Wang},
  \citenamefont {Wen}, \citenamefont {Xiang},\ and\ \citenamefont
  {Zhou}}]{wang2013modeling}%
  \BibitemOpen
  \bibfield  {author} {\bibinfo {author} {\bibfnamefont {Y.}~\bibnamefont
  {Wang}}, \bibinfo {author} {\bibfnamefont {S.}~\bibnamefont {Wen}}, \bibinfo
  {author} {\bibfnamefont {Y.}~\bibnamefont {Xiang}}, \ and\ \bibinfo {author}
  {\bibfnamefont {W.}~\bibnamefont {Zhou}},\ }\href@noop {} {\bibfield
  {journal} {\bibinfo  {journal} {IEEE Communications Surveys \& Tutorials}\
  }\textbf {\bibinfo {volume} {16}},\ \bibinfo {pages} {942} (\bibinfo {year}
  {2013})}\BibitemShut {NoStop}%
\bibitem [{\citenamefont {Mishra}\ and\ \citenamefont
  {Keshri}(2013)}]{mishra2013mathematical}%
  \BibitemOpen
  \bibfield  {author} {\bibinfo {author} {\bibfnamefont {B.~K.}\ \bibnamefont
  {Mishra}}\ and\ \bibinfo {author} {\bibfnamefont {N.}~\bibnamefont
  {Keshri}},\ }\href@noop {} {\bibfield  {journal} {\bibinfo  {journal}
  {Applied Mathematical Modelling}\ }\textbf {\bibinfo {volume} {37}},\
  \bibinfo {pages} {4103} (\bibinfo {year} {2013})}\BibitemShut {NoStop}%
\end{thebibliography}%
\bibliographystyle{apsrev4-1}

\end{document}